%% file: inspiral.tex
\documentclass[twocolumn,superscriptaddress,showpacs,prd,aps,amsmath,amssymb,nofootinbib]{revtex4-1}
\usepackage{natbib}
\usepackage{graphicx,color}
\usepackage{amsmath,amssymb}
\usepackage{verbatim}
\usepackage{float}
\usepackage{wasysym}
\usepackage{amssymb,graphicx}
\usepackage{epsfig}
\usepackage{psfrag}
\usepackage{dsfont}
\usepackage{amsfonts}
\usepackage{mathrsfs}
\usepackage{multirow}
\usepackage{times}
\usepackage{bm}
\usepackage{hyperref}
\usepackage{xspace}
\usepackage{scalefnt}
\hypersetup{
  colorlinks=true,        
  linkcolor=blue,         
  citecolor=cyan,         
}
\input kigou.tex

\newcommand{\GA}{\alpha}
\newcommand{\GB}{\beta}

\newcommand{\GE}{\epsilon}

\newcommand{\GR}{\rho}

\newcommand{\GX}{\chi}
\newcommand{\GO}{\omega}

\newcommand{\GP}{\phi}

\newcommand{\GU}{\theta}

\newcommand{\be}{\begin{equation}}
\newcommand{\ee}{\end{equation}}

\def\QEQ{{%
    \setbox0\hbox{$I$}%
    \rlap{\hbox to \wd0{\hss--\hss}}\box0
}}

\begin{document}
\title{Effect of spin on the inspiral of binary neutron stars.}

\author{Antonios Tsokaros}
\affiliation{Department of Physics, University of Illinois at Urbana-Champaign, Urbana, IL 61801}
\email{tsokaros@illinois.edu}
\author{Milton Ruiz}
\affiliation{Department of Physics, University of Illinois at Urbana-Champaign, Urbana, IL 61801}
\author{Vasileios Paschalidis}
\affiliation{Departments of Astronomy and Physics, University of Arizona, Tucson, AZ 85719}
\author{Stuart L. Shapiro}
\affiliation{Department of Physics, University of Illinois at Urbana-Champaign, Urbana, IL 61801}
\affiliation{Department of Astronomy \& NCSA, University of Illinois at Urbana-Champaign, Urbana, IL 61801}
\author{K\=oji Ury\=u}
\affiliation{Department of Physics, University of the Ryukyus, Senbaru, Nishihara, Okinawa 903-0213, Japan}

\date{\today}

\begin{abstract}
We perform long-term simulations of spinning binary neutron stars,
with our highest dimensionless spin being $\chi \sim 0.32$. To assess
the importance of spin during the inspiral we vary the spin, and also
use two equations of state, one that consists of plain nuclear matter
and produces compact stars (SLy), and a hybrid one that contains both
nuclear and quark matter and leads to larger stars (ALF2). Using high
resolution that has grid spacing $\Delta x\sim 98$ m on the finest
refinement level, we find that the effects of spin in the phase
evolution of a binary system can be larger than the one that comes
from tidal forces. Our calculations demonstrate explicitly that
although tidal effects are dominant for small spins ($\lesssim 0.1$),
this is no longer true when the spins are larger, but still much
smaller than the Keplerian limit.
\end{abstract}

\maketitle

\section{Introduction}
\label{sec:intro}

The 2017 discovery of a binary neutron star (NS) merger by Advanced LIGO and
Advanced Virgo \cite{Aasi_2015,Acernese_2014}, marked a ``golden moment'' in the
era of multimessenger astronomy, since for the first time a gravitational wave
(GW) signal from a merging binary system that included matter was detected at
the same time as its electromagnetic (EM) counterpart
\cite{TheLIGOScientific:2017qsa,GBM:2017lvd, Monitor:2017mdv,
Chornock:2017sdf,2017GCN.21520....1V,Savchenko:2017ffs}.  Although a neutron
star-black hole system was not ruled out completely \cite{Hinderer:2018pei}, 
the measured individual
masses suggested that GW170817 was more likely produced by a binary NS system,
without excluding more exotic objects \cite{TheLIGOScientific:2017qsa}.

One outstanding problem in current astrophysics is the determination
of the equation of state at supranuclear densities, like the ones
present in binary NS systems
\cite{Lattimer:2004pg,Lattimer:2012nd,Ozel:2016oaf,Baiotti:2016qnr,Paschalidis:2016vmz}.
To tackle this problem one needs to measure accurately the masses and
radii of the component stars \cite{Cutler:1994ys} .  For a binary
system like the one that produced the event GW170817, although the
chirp mass is accurately determined, the degeneracy between the mass
ratio of the component objects and their spins along the orbital
angular momentum, prevents the precise measurement of their individual
masses or the total mass of the system. Also for the radii extraction,
the most promising method is based on the measurement of the tidal
deformability parameter \cite{1994ApJ...420..811L,PhysRevD.77.021502,
  PhysRevD.83.084051,Hinderer:2009ca,PhysRevD.85.124034,PhysRevD.88.044042}. Tidal
effects become important at the end of the inspiral (for GW
frequencies $f_{\rm gw}>500\ {\rm Hz}$ where LIGO sensitivity is
decreased), and they depend on the masses, the equation of state, and
likely the spins of the component objects.

Although the magnitude of spin in binary NS systems is largely
unknown, it is important to realize that since discoveries are based
on the identification of the acquired waveform with a corresponding
one from a bank of templates, failing to incorporate waveforms of
spinning binary NSs will result in a possible reduction or
misinterpretation of observations in those cases where such systems
are realized. Thus, although there is the expectation that any initial
spin a NS exhibits at the moment of its genesis will decay by the time
it enters the LIGO band \cite{lrr-2008-8}, the unbiased approach is to
anticipate the physics of a spinning binary in order to maximize our
potential discoveries \cite{Harry2018}.  On the other hand, given the
fact that the number of the currently known binary NS systems is very
small compared to isolated ones, it is not difficult to expect that
there should exist binary NSs with significant rotation. For a NS in
isolation its rotational frequency, has been observed to be as high as
$f_{\rm max}=716\;{\rm Hz}$, corresponding to a period of $1.4\;{\rm
  ms}$ for PSR J1748-2446ad \cite{2006Sci...311.1901H}.  Assuming a
mass of $m\sim 1.36\ M_\odot$ and a moment of inertia $I\sim 1.1\times
10^{45}\ {\rm gr\; cm^2}$, this yields a dimensionless spin of
$\GX\sim I \GO_{\rm max}/m^2 (c/G)\approx 0.3$.

For the 18 currently known binary NS systems in the Galaxy
\cite{Tauris:2017omb, PhysRevD.98.043002}, the rotational frequencies are
typically smaller. The NS in the system J1807-2500B has a period of $4.2\;{\rm
ms}$, while systems J1946+2052 \cite{Stovall:2018ouw}, and J1757-1854
\cite{Cameron:2017ody}, J0737-3039A \cite{Kramer:2006nb} have periods $16.96,\
21.50,$ and $22.70\;{\rm ms}$, respectively. According to Ref.
\cite{PhysRevD.98.043002}, the periods of these systems at merger will be
$18.23,\ 27.09$, and $27.17\;{\rm ms}$, respectively.  When one performs
numerical relativistic simulations and tries to do accurate GW analysis one
cannot model these binaries as irrotational (something that is done in the
majority of simulations), and the spin of each NS must be taken into account. 

In order to perform a constraint-satisfying evolution of spinning binary NSs,
initial data that incorporate spin must be constructed. A self-consistent
formulation for such disequilibrium was first presented in \cite{Tichy:2011gw,
2012PhRvD..86f4024T} using the pseudospectral \sgrid code and the first
evolutions of the last orbits before merger in \cite{Bernuzzi:2013rza} using the
\bam code. The authors found that accurate GW modeling of the merger requires
the inclusion of spin, even for moderate magnitudes expected in binary NS
systems. First evolutions of self-consistent binary NS initial data with spins
in arbitrary directions were presented in 
\cite{PhysRevD.92.124012} where also eccentricity-reduced techniques where
successfully implemented.

Long-term binary NS evolutions geared towards precise GW waveform construction
where pursued by several groups (see \cite{Baiotti:2016qnr} for a recent
review).  The most accurate of them used non-spinning initial data and tracked
the binaries for more than 15 orbits with a subradian-order error
\cite{PhysRevD.96.084060}. The authors used high resolution ($\Delta
x\approx63-86\ {\rm m}$ inside the NSs) together with eccentricity-reduced
initial data. For such high resolutions they found that the phase error in the
GW is $\sim 0.1$ rad among a total phase of $\gtrsim 210$ rad. On the other hand
they report that even with a small residual eccentricity, of the order of $\sim
10^{-3}$, it is still difficult to get accurate quasicircular waveforms.
Accurate models of GWs from irrotational binary NSs studying tidal effects were constructed in 
\cite{Baiotti:2008ra,Baiotti:2010xh,PhysRevD.87.044001,PhysRevD.91.064060,PhysRevD.93.064082,PhysRevD.97.044044}.
The longest irrotational binary NS simulations where presented in
\cite{PhysRevD.93.124062} where for the first time more than 22 orbits where
tracked for a $\Gamma=2$ polytropic EoS using the \spec code.

On the other hand spinning binary NS systems have been examined in detail in 
Refs. \cite{PhysRevD.95.044045, PhysRevD.97.064002} with all possible configurations 
of aligned and misaligned spin as well as with unequal masses. A high resolution
study was presented in Ref. \cite{Dietrich:2018upm}. For
dimensionless spin magnitudes of $\chi \sim 0.1$ the authors found that both
spin-orbit interactions and spin induced quadrupole deformations affect the
late-inspiral dynamics, which however is dominated by tidal effects
(approximately 4 times larger). Closed-form tidal approximants for GWs have been 
presented in Refs. \cite{PhysRevD.97.044044,PhysRevD.96.121501} .
For other dynamical spacetime simulations with spinning binary NSs see
also \cite{Bernuzzi:2013rza,Kastaun2013,Kastaun2015,Tacik:2015tja,
bauswein2015exploring,Dietrich:2015pxa,EPP2015,PEPS2015,EPPS2016,East:2016zvv,
Dietrich:2017xqb,Ruiz:2019ezy,Most:2019pac}.

In this paper we use the \illinois code to compare the GW of a long
inspiral coming from an irrorational binary NS with a highly spinning
one. The initial spinning configurations have been constructed with
the \cocal code \cite{Tsokaros:2015fea,Tsokaros:2018dqs} whose
accuracy has been tested extensively \cite{Tsokaros:2016eik}, and has
been used to evolve one of the highest spinning binary NSs to date
\cite{Ruiz:2019ezy}.  The simulations performed here are the longest
using the \illinois code and they provide a benchmark in order to go
to larger orbital separations, and to construct reliable waveforms.
We use two piecewise polytropic equations of state (EoS) and a high
spin ($\chi \sim 0.32$ for one binary configuration) to assess its
influence in the latest $\sim 12-17$ orbits before merger. We find
that although tidal terms dominate when the NS spins are small, this
is no longer true for higher spins.  This is in qualitative accordance
with the post-Newtonian analysis \cite{Harry2018} who found that large
spins could cause significant mismatches.  In our study a soft EoS
(SLy, compact star) with a $\chi \sim 0.2$ spin produced a phase
difference with respect to the irrotational case of $\sim 23$ radians,
while a stiffer EoS (ALF2, larger NS radius) with a $\chi \sim 0.32$
spin, produced $\sim 40$ radians. This phase difference is expected to
be even larger for higher spins and highlights the fact that GW data
analysis will be compromised if spin effects are neglected.

The present study has two main caveats. First, our initial
quasiequilibrium models exhibit residual eccentricity which
contaminates late inspiral waveforms and prevents an accurate GW
analysis. As mentioned in \cite{PhysRevD.96.084060}, even when
eccentricity reduction was implemented, there was still existing
artifacts that necessitated the removal of the first couple of orbits
in the GW analysis.  Currently our initial data solver does not
account for eccentricity.  Second, due to our limited resources we
have not performed a resolution study to test for convergence and
quantify errors. In spite of these caveats, we employ the highest
resolution used to date for highly spinning binary systems with our
finest grids having $\Delta x\sim 98\ {\rm m}$. According to
\cite{PhysRevD.96.084060} employing $\Delta x_{\rm min} \le 100$ m one
achieves sub-radian accuracy ($\sim 0.2$ rad) and nearly convergent
waveforms in approximately 15 orbits.  Finally, we do not test if
there are any outer boundary effects in these simulations.  We plan to
address these shortcomings in the near future.

Here we employ geometric units in which $G=c=M_\odot=1$, unless stated           
otherwise. Greek indices denote spacetime dimensions, while            
latin indices denote spatial ones.

\section{Numerical methods}

The numerical methods used here are those implemented in the \cocal           
and \illinois codes, and have been described in great detail in our            
previous works \cite{Tsokaros:2015fea, Tsokaros:2016eik, Tsokaros:2018dqs,
Uryu:2011ky, Etienne:2012te,UIUC_PAPER1, UIUC_PAPER2,prs15}.
Therefore we will only summarize the most important features here.     
In the following sections we describe our initial configurations, the grids used
in our simulations, the EoSs, and how we compute the GWs.

\subsection{Initial data}
\label{ssec:id}

To probe the effect of spin during the inspiral phase of a merging
binary we evolve irrotational as  well as spinning configurations that are
constructed with our initial data solver \cocal \cite{Uryu:2011ky,
Tsokaros:2015fea, Tsokaros:2016eik, Tsokaros:2018dqs} in order to make a
critical comparison.  The simplest spinning configurations are the so-called
corotating solutions, that were historically the first ones to be computed
\cite{1997PhRvL..79.1182B,1998PhRvD..57.7299B, 1998PhRvD..58j7503M}, and
describe two NSs tidally locked, as the Moon is in the Earth-Moon system. Although
this state of rotation is considered unrealistic since the viscosity in NSs is too
small to achieve synchronization \cite{1992ApJ...400..175B,
1992ApJ...398..234K}, it is still a viable choice to investigate when the
separation (orbital velocity) is large but not extremely large so that the NSs
have a reasonable spin. In this work we consider binaries starting at an orbital
angular velocity $\Omega=6\times 10^{-3}$, which translates into $f=194\ {\rm
Hz}$ for the NS rotation rate. This frequency is well within the realistic
regime of spins for NSs which, as mentioned in the introduction is observed to
be as high as $f_{\rm max}=716\;{\rm Hz}$.  Assuming a spinning binary NS system
is formed with individual NS frequencies at $f=194\ {\rm Hz}$, then from that
point on the corotating state is no longer preserved in a perfect fluid evolution,
and therefore the argument about sychronization is not applicable.

Apart from the corotating solutions we construct generic aligned and
antialigned spinning solutions using the formulation developed by
Tichy \cite{Tichy:2011gw}. Following \cite{Tsokaros:2018dqs} the
calibration of the spin is done with the use of the circulation
concept along an equatorial ring of fluid. The \cocal code can produce
binaries of a prescribed circulation (along with the rest mass and
orbital separation).  Therefore, for each EoS we compute the
corotating binary and measure its circulation $\mathcal{C}_{\rm
  cor}$. Having that value we compute generic spinning binaries whose
circulation is some multiple of the corotating one. In particular,
aligned binaries have a circulation which is approximately
$2\mathcal{C}_{\rm cor}$, while the antialigned binaries
$-\mathcal{C}_{\rm cor}$. Thus our binary systems exhibit a wide range
of spins, which, in addition, fall into the realistic regime of
rotation rates.

Regarding the EoSs in this work we choose the ALF2 \cite{Alford2005}
(a hybrid EoS with mixed APR \cite{Akmal1998a} nuclear matter and
color-flavor-locked quark matter) and the SLy \cite{Douchin01} (pure
hadronic matter) EoSs. A NS with Arnowitt-Deser-Misner (ADM) mass of
$1.4\ M_\odot$ for these EoSs has the characteristics shown in Table
\ref{tab:eos}. The tidal deformability parameter is given by
$\Lambda=2k_2(M/R)^{-5}/3$, with $k_2$ the tidal Love number computed
from linear perturbations of the spherical solution
\cite{Hinderer:2007mb}. As shown in Table \ref{tab:eos} the ALF2 EoS
is stiffer than the SLy EoS, in the sense that it predicts larger
radii for the same gravitational mass, and larger tidal
deformability. The purpose of our work is to understand the importance
of spin on the observed waveforms, therefore our choice of EoS was
dictated on the one hand from the need to explore typical neutron
matter (SLy) as well as more exotic compositions (ALF2), and on the
other hand from current EoS constraints. These two EoSs are broadly
consistent with a number of studies that use the GW170817 event to
constrain the radii and tidal deformabilities of NSs
\cite{Annala2017,Bauswein2017b,Radice2017b,Abbott2018b,Most2018,Kiuchi2019}.

\begin{table}[h]                                                             
\caption{Characteristics of a spherical $M=1.4M_\odot$ NS for the 2 EoSs used in 
this work.} 
\centering                                                                       
\setlength{\tabcolsep}{0.25em}                                                   
\label{tab:eos}                                                                  
\begin{tabular}{|c|c|c|c|c|}                                                     
\hline                                                                           
EOS   & $M^{(a)}$  &  $R$(km)$^{(b)}$  &  $M/R$  &  $\Lambda$  \\                              
\hline                                                                           
ALF2  & 1.40 &  12.39    &  0.1670 &  589.4     \\                               
SLy   & 1.40 &  11.46    &  0.1804 &  306.4     \\                               
\hline                                                                           
\end{tabular}                     
\begin{flushleft}                                                                
  $^{(a)}$ ADM mass.\hspace{2cm}
  $^{(b)}$ Areal radius.          
\end{flushleft}                                                 
\end{table}

\setlength{\tabcolsep}{4pt}
\begin{table*}   
\caption{8 initial data configurations used in this work. The first 4 lines
correspond to antialigned spinning, irrotational, corotating, and aligned
spinning for the ALF2 EoS. Similarly the next 4 lines correspond to the SLy EoS.
All binary sets have ADM mass $M=2.72$, and $\Omega=6\times 10^{-3}$. 
$M_0$ is the rest mass of
each NS, $\GX$ the
dimensionless spin, ${\rm P}$ the NS spin period in milliseconds,
$J$ the ADM angular momentum, $R_x,\ R_y,\ R_z$
the coordinate radii, $\GR_0$ the maximum rest-mass density, and
$\mathcal{C}$ the equatorial circulation.  To convert to cgs units multiply mass, density
and distance by $1.989\times 10^{33}\ {\rm g}$, $6.173\times 10^{17}\ {\rm
g/cm^3}$, and $1.477\times 10^5\ {\rm cm}$, respectively. } 
\label{tab:idruns}                                                                        
\begin{tabular}{l|cccccccccc}                                
\hline                                                                           
\hline                                                                           
Name & Separation & $M_0[M_\odot]$ & $\GX$ & ${\rm P} [{\rm ms}]$ & $J$ & $R_x$ & $R_y$ & $R_z$ & 
     $\rho_0(\times 10^{-3})$ & $\mathcal{C}$  \\
\hline
spALF2-1c & $39.98$ & $1.511$ & $-0.1703$            & $-4.898$            & $7.779$ & $6.791$ & $6.668$ & $6.621$ 
          & $1.043$ & $-2.809$ \\ 
irALF2    & $39.94$ & $1.512$ & $-0.0020$            & N/A                 & $8.176$ & $6.747$ & $6.635$ & $6.661$ 
          & $1.050$ & $\phantom{-}0.000$ \\  
coALF2    & $39.89$ & $1.511$ & $\phantom{-}0.1637$  & $\phantom{-}5.159$  & $8.571$ & $6.785$ & $6.675$ & $6.621$ 
          & $1.043$ & $\phantom{-}2.813$ \\  
spALF2+2c & $40.08$ & $1.510$ & $\phantom{-}0.3206$  & $\phantom{-}2.522$  & $9.054$ & $6.904$ & $6.788$ & $6.513$ 
          & $1.025$ & $\phantom{-}5.618$ \\  
spSLy-1c  & $39.97$ & $1.518$ & $-0.1006$            & $-4.844$            & $7.843$ & $6.161$ & $6.078$ & $6.051$ 
          & $1.403$ & $-2.433$\\  
irSLy     & $39.92$ & $1.519$ & $-0.0016$            & N/A                 & $8.171$ & $6.130$ & $6.051$ & $6.074$ 
          & $1.408$ & $\phantom{-}0.000$   \\
coSly     & $39.88$ & $1.519$ & $\phantom{-}0.0982$  & $\phantom{-}5.159$  & $8.503$ & $6.158$ & $6.081$ & $6.050$ 
          & $1.403$ & $\phantom{-}2.436$ \\  
spSLy+2c  & $40.05$ & $1.517$ & $\phantom{-}0.1805$  & $\phantom{-}2.481$  & $8.906$ & $6.247$ & $6.168$ & $5.981$ 
          & $1.387$ & $\phantom{-}4.866$ \\  
\hline
\hline                                                                           
\end{tabular}                                                                                                                      
\end{table*}

In Table \ref{tab:idruns} we report the 8 initial configurations we
consider in this work.  We fix the ADM mass of the binary systems to
be $M=2.72$ and their orbital angular velocity at $\Omega=6\times
10^{-3}$.  For the ALF2 EoS, a spherical isolated NS with ADM mass
$1.36$\footnote{The maximum spherical ADM mass for the ALF2 EoS is
  $1.99\ M_\odot$ and the maximum compactness $0.26$, while for the
  SLy EoS the corresponding values are $2.06\ M_\odot,\ 0.33$, respectively.} has
compactness $0.1625$, tidal Love number $k_2=0.1191$, and tidal
deformability $\Lambda=701.3$. For the SLy EoS with the same spherical
mass ($1.36$) we have a higher compactness $0.1752$, smaller tidal
Love number $k_2=0.09298$, and smaller tidal deformability parameter
$\Lambda=371.2$.  Following the argument of the previous paragraph we
notice that the most extreme dimensionless spins $\GX \equiv J_{\rm
  ql}/(M_{\rm ADM}/2)^2$ (here $J_{\rm ql}$ refers to the quasilocal
angular momentum \cite{Tsokaros:2018dqs}) happen in the ALF2 EoS
($-0.1703$ and $0.3206$), which are the highest evolved for a period
of 16 orbits.  In the maximum spin case the quasilocal spin is $\sim
6.5\%$ of the ADM angular momentum of the system. The spin period of
each NS is computed as ${\rm P}=2\pi/\Omega_s^z$ where $\Omega_s^z$
the parameter that controls the spin of the NS
\cite{Tsokaros:2018dqs}. This is an approximate measure of the
rotation period of the NS not rigorously defined in general
relativity, except in the corotational case.

\begin{table*}
\caption{Grid parameters used for the evolution of each  
binary configuration of Table \ref{tab:idruns}. The computational grid consists
of three sets of eight nested refinement boxes, the innermost ones centered on
each star and on the origin of the computational domain.
Parameter $\Delta x_{\rm max}$ 
is the step interval in the coarser level, while $\Delta x_{\rm min}$ in the
finer. To convert to physical units multiply by $1.477\ {\rm km}.$}            
\label{tab:grid}                 
\begin{tabular}{ccccccccc}                                                      
\hline                                                                           
\hline                                                                           
$x_{\rm min}$ & $x_{\rm max}$ & $y_{\rm min}$ & $y_{\rm max}$ & $z_{\rm min}$ & $z_{\rm max}$ & 
Grid hierarchy (Box half-length) & $\Delta x_{\rm max}$ & $\Delta x_{\rm min}$  \\                                                  
\hline                                                                           
$-1024$ & $1024$ &  $-1024$ &  $1024$ &  $0$ &  $1024$ & $\{8.0,16.0,32.0,64.0,128.0,256.0,512.0,1024.0\}$ & 
$8.5\bar{3}\ \ \ $ & $0.0\bar{6}$  \\   
\hline                                                                           
\hline                                                                           
\end{tabular}                                                                    
\end{table*}

\subsection{Evolution}
\label{ssec:evolution}

We use the \illinois adaptive-mesh-refinement code that has been
embedded in the \cactus/\carpet infrastructure
\cite{AllAngFos01,cactusweb,Carpet,carpetweb}, and employs the
Baumgarte--Shapiro--Shibata--Nakamura (BSSN) formulation of the
Einstein's equations~\cite{shibnak95,BS} (for a detailed discussion
see also~\cite{BSBook}) to evolve the spacetime and matter
fields. Fourth order, centered finite differences are used for spatial
derivatives, except on shift advection terms, where we employ fourth
order upwind differencing. Outgoing wave-like boundary conditions are
applied to all BSSN evolved variables.  These variables are evolved
using the equations of motion (9)-(13) in~\cite{Etienne:2007jg}, along
with the $1+$log time slicing for the lapse $\alpha$ and the
``Gamma--freezing" condition for the shift $\beta^i$ cast in first
order form (see~Eq.~(2)-(4) in~\cite{Etienne:2007jg}). For numerical
stability, we set the damping parameter $\eta$ appearing in the shift
condition to $\eta=2.312/M$.  For further stability we modify the
equation of motion of the conformal factor $\phi$ by adding a
constraint-damping term (see Eq.~19 in~\cite{DMSB}) which damps the
Hamiltonian constraint. We set the constraint damping parameter to
$c_H=0.08$.  Time integration is performed via the method of lines
using a fourth-order accurate Runge-Kutta integration scheme with a
Courant-Friedrichs-Lewy (CFL) factor set to $0.5$.  We use the Carpet
infrastructure~\cite{Carpet,carpetweb} to implement moving-box
adaptive mesh refinement, and add fifth order Kreiss-Oliger
dissipation~\cite{goddard06} to spacetime and gauge field variables.
                                                                                 
The equations of hydrodynamics are solved in conservation-law form adopting the
high-resolution shock-capturing methods described in \cite{Etienne:2011re, Etienne:2010ui}.  The               
primitive, hydrodynamic matter variables are the rest mass density,              
$\GR_0$, the pressure $P$ and the coordinate three velocity                      
$v^i=u^i/u^0$. The specific enthalpy is written as $h=1+\GE+P/\GR_0$, and                 
therefore the stress energy tensor is $T_{\GA\GB}=\GR_0 hu_\GA u_\GB+            
P g_{\GA\GB}$. Here, $\GE$ is the specific internal                              
energy.                                                                                                                                        
To close the system an EoS needs to be provided and for that                     
we follow \cite{2011PhRvD..84j4032P,2011PhRvD..83f4002P} 
where the pressure is decomposed as a             
sum of a cold and a thermal part,                                                
\be                                                                              
P = P_{\rm cold} + P_{\rm th} = P_{\rm cold} + (\Gamma_{\rm th}-1)\GR_0 (\GE-\GE_{\rm cold})
\label{eq:pre}                                                                   
\ee 
where 
\be \GE_{\rm cold} = -\int P_{\rm cold} d(1/\GR_0)                     
=\frac{k}{\Gamma-1}\GR_0^{\Gamma-1} + const.\ .  
\ee 
Here $k,\Gamma$             
are the polytropic constant and exponent of the cold part (same as the
initial data EoS) and $\Gamma_{\rm th}=5/3$~\cite{2011PhRvD..84j4032P}.
The constant that appears in the formula above (which is zero for a
single polytrope), is fixed by the continuity of pressure at the dividing
densities between the different pieces of the piecewise polytropic
representation of the ALF2 and SLy EoSs.

The grid hierarchy used in our simulations is summarized in Table
\ref{tab:grid}. It consists of three sets of nested mesh refinement
boxes, two of them centered on the locations of the two density maxima
on the grid (the ``centers'' of the NSs), and the third one at the
origin of the computational domain $[-1024,1024]^2\times[0,1024]$.
For each case listed in Table \ref{tab:idruns} halving the value under
``Separation'' column provides the initial coordinate location of the
centers of the NSs (one is on the positive x-axis and the other on the
negative x-axis), which is the coordinate onto which two of our nested
refinement levels are centered on. Each nested set consists of eight
boxes that differ in size and in resolution by factors of two. The
half-side length of the finest box (which in our case is $8.0$) is
covered by 120 points which results in $\Delta x_{\rm min}\sim
8.5\bar{3}/2^7=0.0\bar{6}\approx 98\ {\rm m}$.  The half-side length
of the finest box is chosen according to the initial neutron star
equatorial radius $R_x$, and typically is $1.2-1.3$ times $R_x$.  This
means that the neutron star radius is initially covered by 92 to 104
points.  Reflection symmetry is imposed across the orbital plane.

In comparison with other works our resolution is 2.5 finer than the
highest resolution used in \cite{Dietrich:2016lyp} and slightly higher
than the high resolution spinning runs in
Ref. \cite{Dietrich:2018upm}.  According to \cite{PhysRevD.96.084060}
that has presented the most accurate gravitational waveforms for
irrotational binaries to date, one needs $\Delta x_{\rm min} \le 100$
m to achieve sub-radian accuracy ($\sim 0.2$ rad) and nearly
convergent waveforms in approximately 15 orbits. Although we did not
resolution study, we used a very high resolution in order to fulfill
the requirement of Ref.  \cite{PhysRevD.96.084060}.

\begin{figure}                                                                   
\begin{center}                                                                   
\includegraphics[width=0.99\columnwidth]{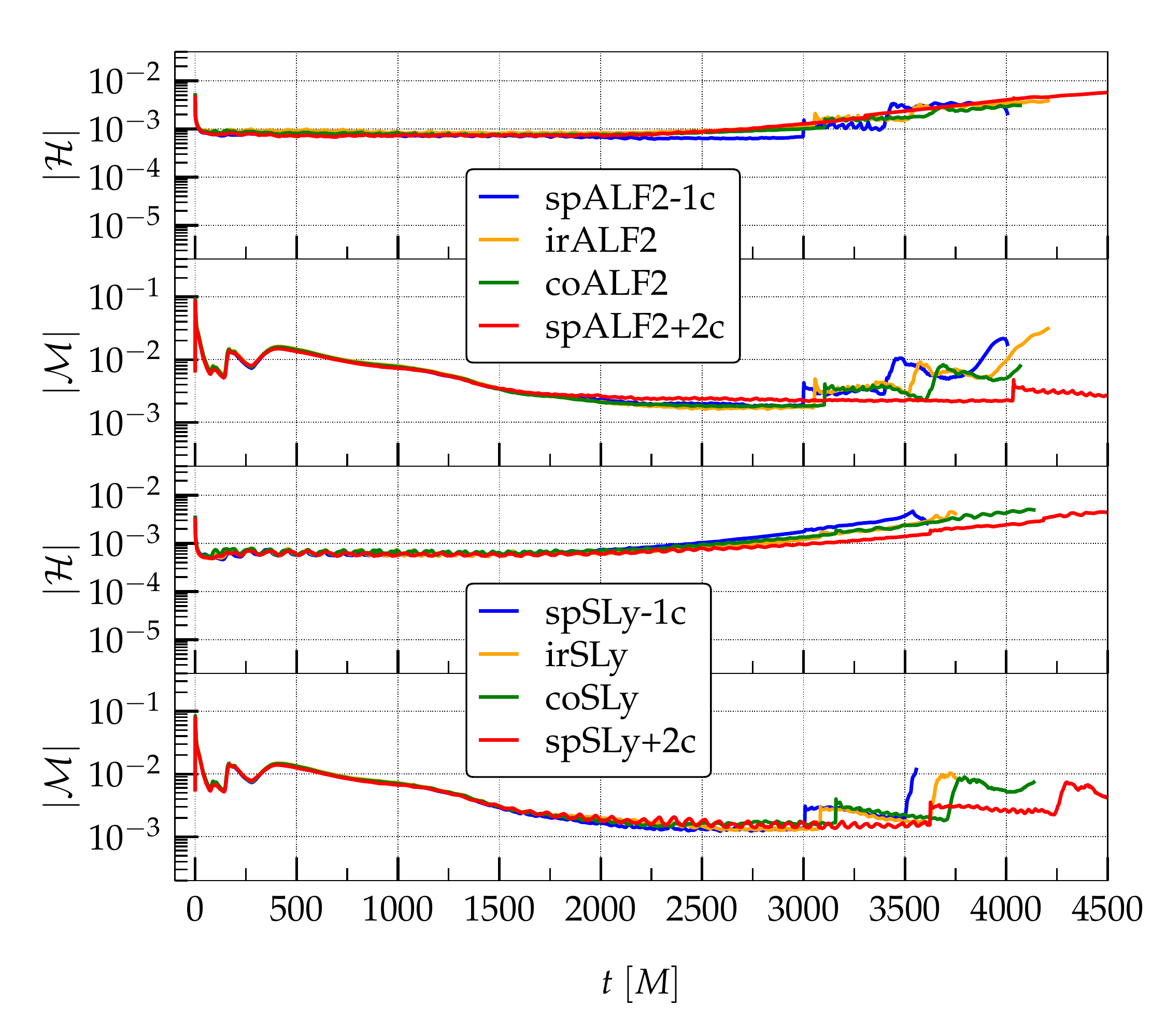}                       
\caption{Top two panels show the Hamiltonian and momentum violations for all models
of the ALF2 EoS. Bottom two panels similarly for the SLy EoS.}                                                           
\label{fig:con}                                                               
\end{center}                                                                     
\end{figure}

In Fig. \ref{fig:con} we plot the constraint violations  for all models using
the diagnostics of Ref. \cite{Etienne:2007jg}.  Models spALF2-1c, spSLy-1c,
irSLy collapse promptly to a black hole upon merger, while the others lead to
hypermassive NSs. As we can see the violations coming from the spinning cases
are identical to those of the irrotational or corotating ones.  The magnitude of
the violations differs from those reported in \cite{Tsokaros:2016eik} since the
\illinois code uses different normalization factors than the \whiskythc code
\cite{Radice:2013hxh,Radice2013c}.

\begin{figure*}
\begin{center}                                                                   
\includegraphics[width=0.98\columnwidth]{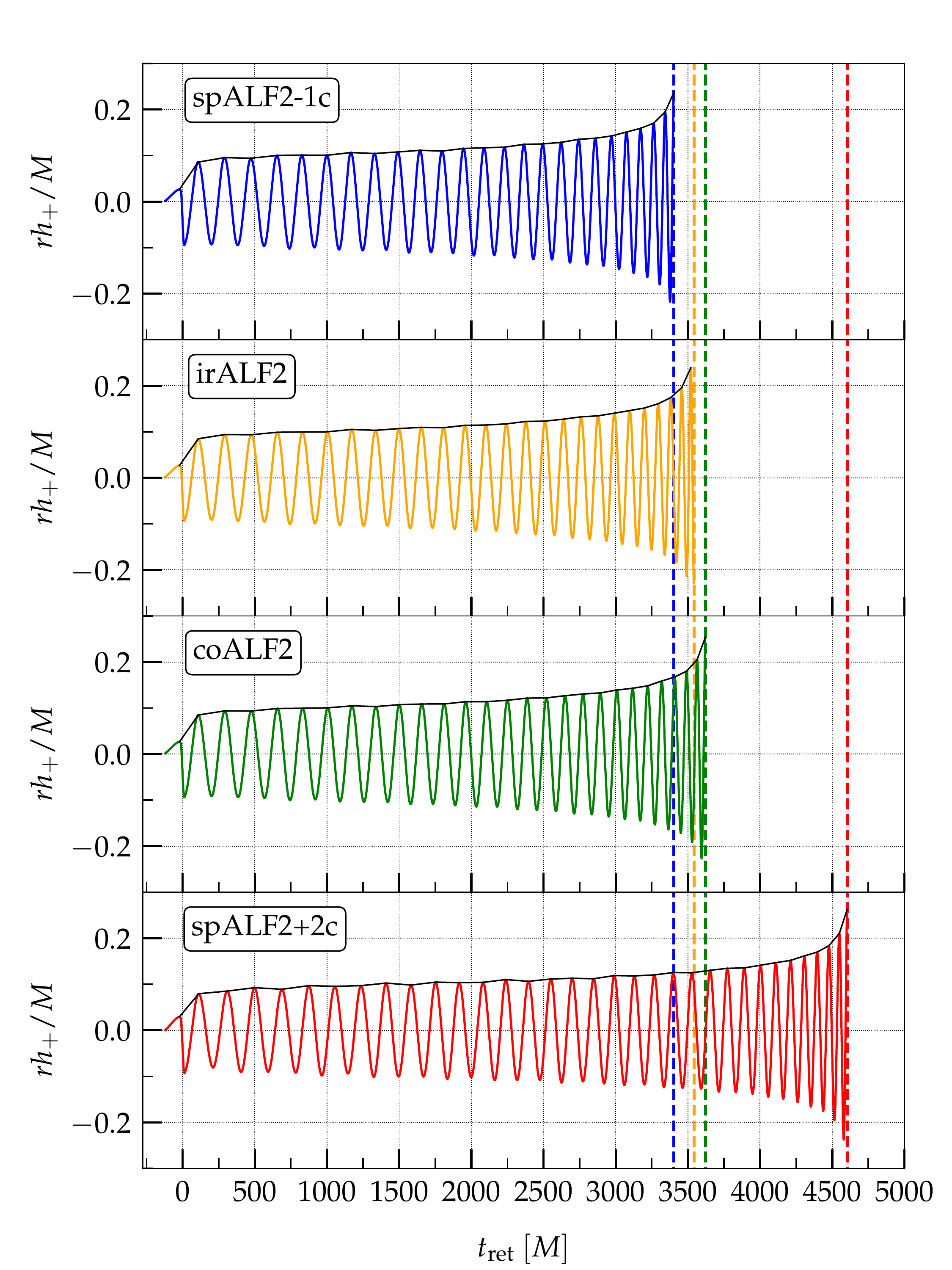}                       
\includegraphics[width=0.98\columnwidth]{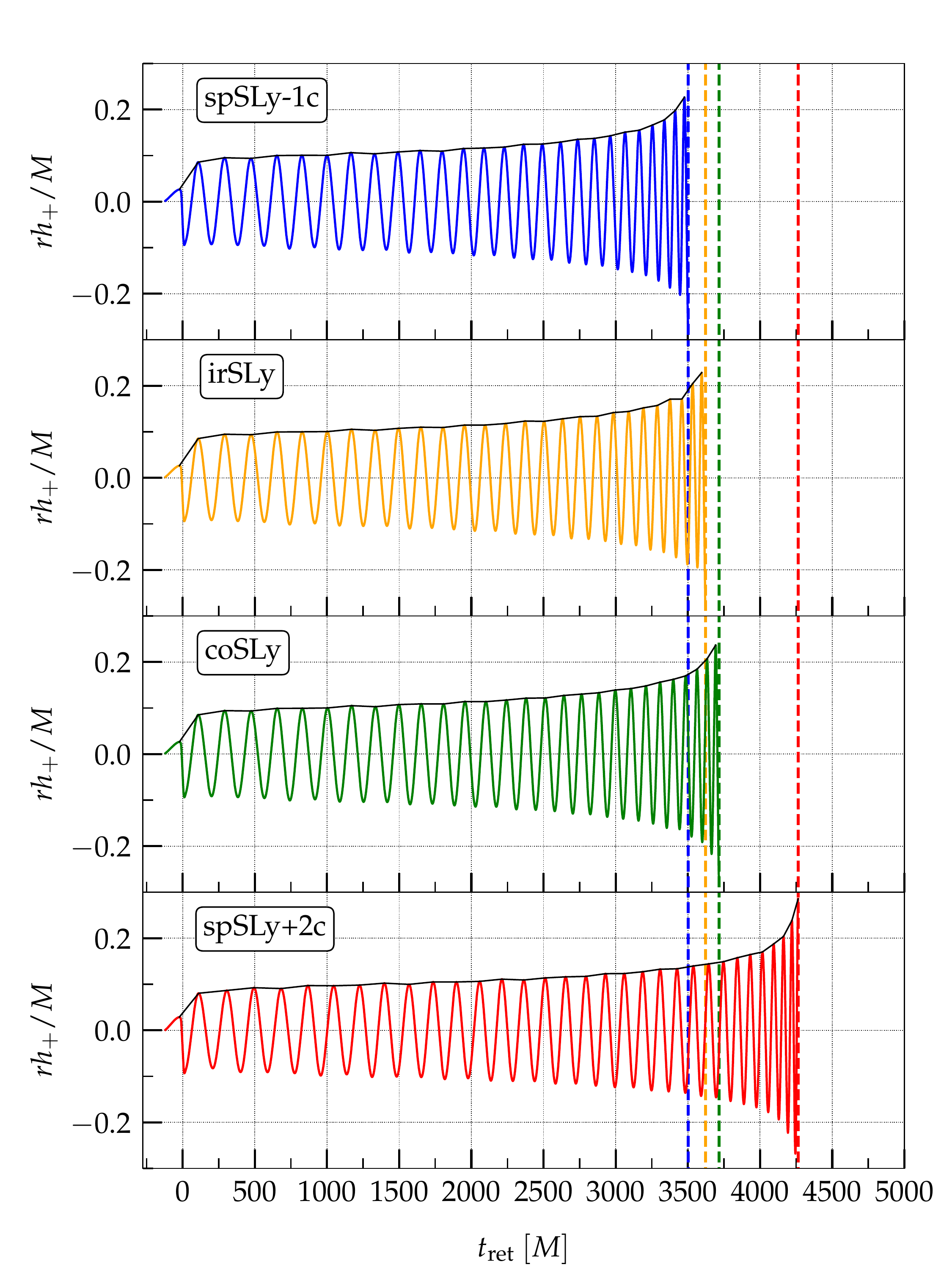}                       
\caption{The strain of the plus polarization of the (2,2) GW mode for the ALF2 (left
column) and the SLy (right column) EoSs. From top to bottom the binaries 
correspond to antialigned spinning, irrotational, corotating, and aligned 
spinning, respectively. Dashed vertical lines denote the time of the maximum 
amplitude $h=\sqrt{h_{+}^2+h_{\times}^2}$.}                                                           
\label{fig:rhoM}                                                               
\end{center}                                                                     
\end{figure*}

\subsection{GW extraction}
\label{ssec:gwex}

Extraction of GWs is performed using the complex Weyl scalar $\Psi_4$
and the fact that $\Psi_4=\ddot{h}_{+} - i \ddot{h}_{\times}$
\cite{BSBook, Maggiore2007, Ruiz:2007yx}.  Expanding in terms of the
spin-weighted spherical harmonics with spin weight $-2$ \be
\Psi_4(t,r,\GU,\GP) = \sum_{\ell=2}^{\infty} \sum_{m=-\ell}^{\ell}
\Psi_4^{\ell m}(t,r)\ _{-2}Y_{\ell m}(\GU,\GP)
\label{eq:psi4}  
\ee
and the strain $h=h_{+}-ih_{\times} $ of the GW will be 
\be
h(t,r,\GU,\GP) = \int_{-\infty}^t dt' \int_{-\infty}^{t'} dt''\
\Psi_4(t'',r,\GU,\GP) \ .
\label{eq:h}
\ee For the 8 simulations performed here (with outer boundary at
$x=y=z=1024$) we extract the GW coefficients $\Psi_4^{\ell m}(t,r)$ at
seven radii, $R_{\rm gw}\in\{120,240,300,460,600,720,840\}$, in order
to make sure that we have a waveform converged with radius. These
coefficients are then expressed in terms of the retarded time $t_{\rm
  ret}=t-r_\star$ where $r_\star = r_A +2M\ln\left(r_A/(2M)-1\right)$
is the so-called tortoise coordinate.  Here $r_A=\sqrt{A_{\rm
    gw}/(4\pi)}$ is the areal (Schwarzschild) coordinate and $A_{\rm
  gw}$ the proper area of a coordinate sphere of radius $R_{\rm gw}$.

\begin{figure*}
\begin{center}                                                                   
\includegraphics[width=2.05\columnwidth]{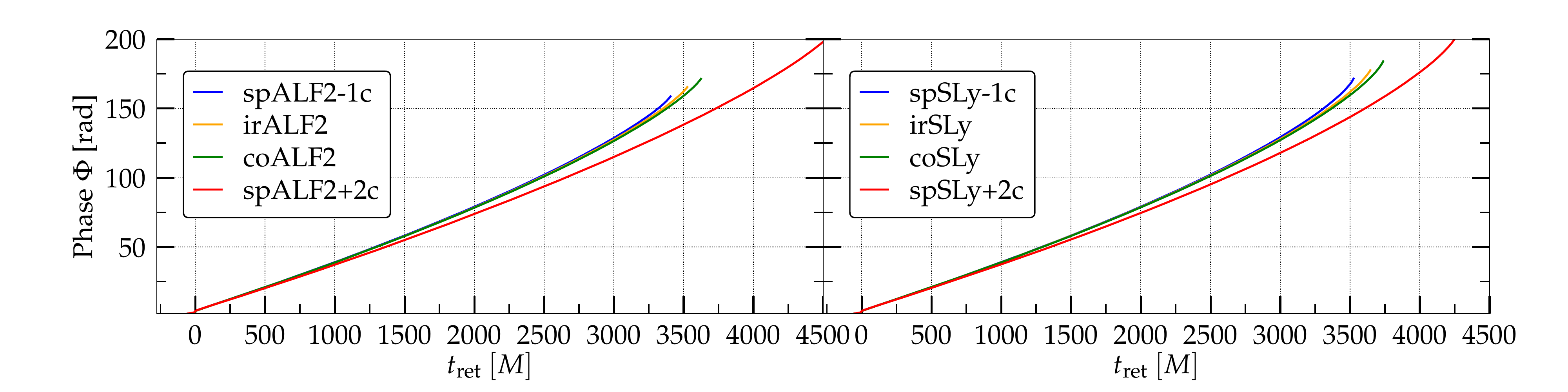}                       
\includegraphics[width=2.05\columnwidth]{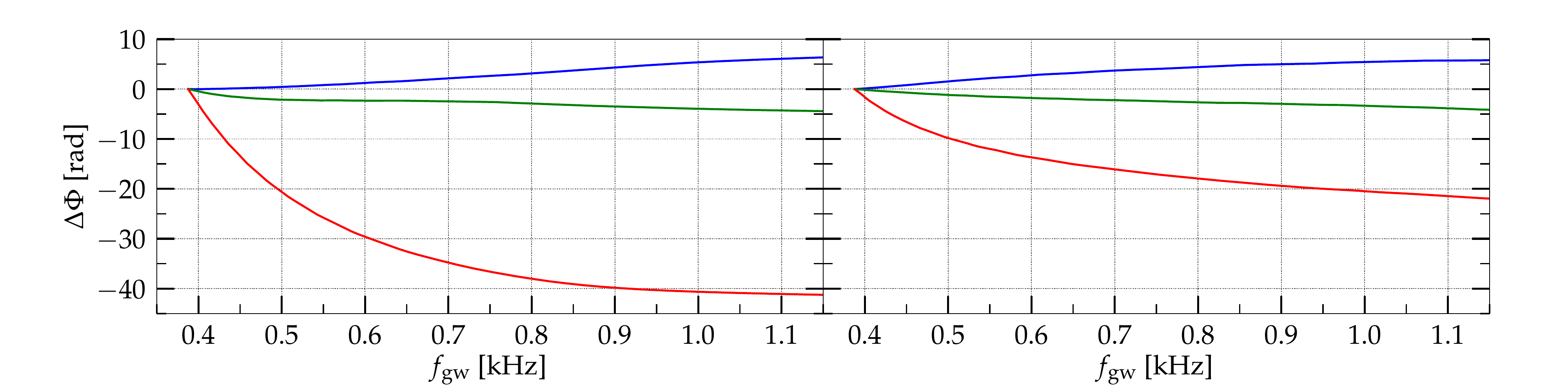}                       
\caption{Left (right) column ALF2 (SLy) EoS. Top panels shows the phase evolution 
with respect to the retarded time while the bottom panels the phase difference between 
the irrotational and spinning models vs the GW frequency of the $(2,2)$ mode.}                                                           
\label{fig:phase}                                                               
\end{center}                                                                     
\end{figure*}

\begin{figure}
\begin{center}                                                                   
\includegraphics[width=0.98\columnwidth]{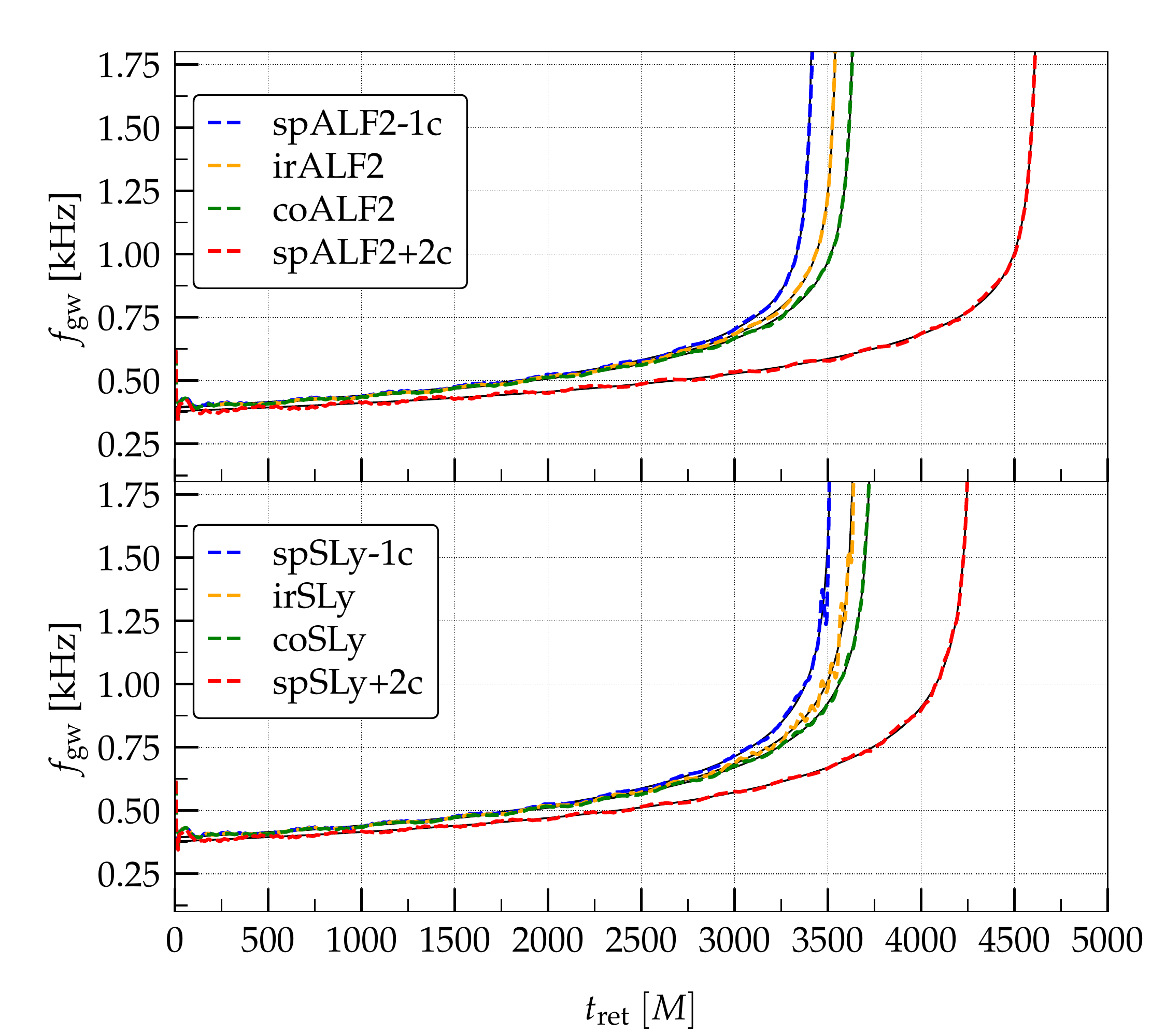}                       
\caption{The gravitational wave frequency for the $l=2$, $m=2$ mode with respect
to the retarded time for both the ALF2 (top panel) and SLy (bottom panel) EoSs.
The black curves running through and practically overlapping with the colored ones 
are the fits using Eq. \eqref{eq:fgwfit}. }                                                           
\label{fig:omega22}                                                               
\end{center}                                                                     
\end{figure}

In order to calculate the strain, Eq. (\ref{eq:h}), we have to perform the 
double time integrations of the coefficients $\Psi_4^{\ell m}(t,r)$ and for that we
follow the recipe of Ref. \cite{Reisswig:2010di} which strongly reduces spurious
secular nonlinear drifts of the waveforms. First the Fourier transform
$\Psi_4^{\ell m}(\GO,r)$ of $\Psi_4^{\ell m}(t,r)$  is calculated and then
the strain coefficients are computed according to 
\be
h^{\ell m}(t,r) = -\frac{1}{2\pi}\int_{-\infty}^{+\infty} \frac{\Psi_4^{\ell
m}(\GO,r)}{{\rm max}(\GO,\GO_0)^2}e^{i\GO t} d\GO  \ .
\label{eq:hft}
\ee
We choose $\GO_0=\Omega (t=0)$. 
Since in this work we simulate equal mass binaries and we are interested in the
inspiral phase (up to merger) of identical stars, we will focus only at the $(\ell,m)=(2,2)$ mode.
From now on we will denote this GW mode by $h=h_{+}^{22}-ih_{\times}^{22}$ and $\Phi$ the 
phase of $h$ at a specific radius, therefore we will write
\be
 h(t) = A(t) e^{i\Phi (t)}\ . 
\label{eq:h1}
\ee
The GW angular frequency is defined as 
\be
\Omega_{\rm gw} = 2\pi f_{\rm gw} = \frac{d\Phi}{dt} \ .
\label{eq:gwo}
\ee

\section{Results}

In Fig. \ref{fig:rhoM} we plot the real part of the gravitational wave
strain $h$ vs the retarded time for the 8 simulations of Table
\ref{tab:idruns}. The left column corresponds to the ALF2 models while
the right to the SLy ones. From top to bottom we plot the spinning
binaries with their spin antialigned with the orbital angular
momentum, the irrotational, the corotating, and the aligned spinning
ones. All waveforms are terminated at their peak amplitude (peak of
$h=\sqrt{h_{+}^2+h_{\times}^2}$) that corresponds to the merger of the
two neutron stars. The time of the peak amplitude of $h$ is not
identical with the time of the peak amplitude of $h_{+}$, but very
close to it\footnote{In Fig.  \ref{fig:rhoM} these two times are
  indistinguishable and essentially coincide with the dashed vertical
  lines.}. The so-called hang-up effect \cite{clz06}, which was
identified in BNS simulations
\cite{Dietrich:2015pxa,Tsatsin2013,Kastaun2013,Dietrich:2016lyp,Ruiz:2019ezy}
is clear in these waveforms.
Comparing the irrotational waveforms of the two EoSs we see that the
ALF2 binary merges earlier than the SLy one, in agreement with the
fact that the tidal deformability of ALF2 is larger than the SLy one
(see Table \ref{tab:peaks} for exact merger times and
frequencies\footnote{For the irrotational cases the frequencies are in
  agreement with $t_{\rm mrg}-\Lambda$ relations reported in
  \cite{Read:2013zra,Bernuzzi:2014kca,Takami:2014tva}, where $t_{\rm
    mrg}$ marks the time of the peak amplitude $h$.}).  Among the
corotating models, the ALF2 merges earlier than the SLy even though
its spin is much larger ($0.16$ vs $0.098$) implying that the
combination of tidal effects and the larger orbital separation at
merger (in the ALF2 case) dominate over spin effects. However, the
model spALF2+2c merges later than spSLy+2c; therefore here the much
higher spin of spALF2+2c overcomes the tidal interactions.

The effect of spin can be seen most clearly in Fig. \ref{fig:phase}
where the phase evolution of the gravitational wave signal is plotted
vs the retarded time (top panels). At any given time the slope of the
curves decreases with increasing aligned spin, with the steepest slope
corresponding to the antialigned models (spALF2-1c, spSLy-1c) and the
smaller for the aligned cases (spALF2+2c,spSLy+2c). A steeper phase
slope (antialigned spins) leads to more bound systems, faster phase
evolution, and thus earlier merger \cite{Damour:2001tu}.  In the
bottom panels of Fig. \ref{fig:phase} we plot the phase difference
between the irrotational models and the spinning ones, $\Delta\Phi =
\Phi_{\rm irrot} - \Phi_{\rm spin}$, vs $f_{\rm gw}$, the
gravitational wave frequency of the (2,2) mode, in the LIGO band. The
transition from retarded time to frequency has been accomplished using
the relations shown in Fig. \ref{fig:omega22}. Color lines represent
raw data which exhibit a slight oscillatory behavior that is
characteristic of the presence of eccentricity in the initial
data. More accurate future evolutions will improve this artifact. In
order to remove this residual eccentricity we perform fittings
inspired by the post-Newtonian formalism~\cite{Blanchet:2013haa},
\be 2\pi f_{\rm gw} =
\frac{1}{20M}z^3(c_0 + c_2z^2 +c_3z^3 + c_4z^4)
\label{eq:fgwfit}
\ee
where $z=[(t_c-t)/(20M)]^{-1/8}$ and $t_c$ the coalescence time (maximum
amplitude of the strain). The fitted curves (black lines in \ref{fig:omega22}
that essentially coincide with the colored ones) are used in Fig.
\ref{fig:phase}.  
By direct comparison of the two panels in the bottom row of Fig. \ref{fig:phase}
one can see that for small spins (antialigned (blue) and corotating (green)) 
the two EoSs yield small differences with respect to the irrotational case.
For higher spins significant deviations from the irrotational models appear. In
the post-Newtonian approximation one can identify the magnitude of the
contributions due to different mechanisms, and to lowest order one can calculate
the point particle (like a binary black hole), tidal, spin-orbit, spin-spin from
self-interactions, and spin-spin from mutual interactions
\cite{Dietrich:2016lyp}.  In our case we find that, 
although small spins (depending also on the EoS) result to phase differences of the
order of $\sim 5$ radians (in accordance with Ref. \cite{Dietrich:2016lyp}), 
higher spins, can produce phase differences as large as $\sim 40$ radians within the 
$1\ {\rm KHz}$ band which \textit{are much larger than
the tidal effects}. 

\setlength{\tabcolsep}{6pt}
\begin{table}   
\caption{Retarded time of the peak of the 2,2 mode as well as the corresponding
frequencies for the 8 initial data configurations used in this work. } 
\label{tab:peaks}                                                                        
\begin{tabular}{l|cc}                                
\hline                                                                           
\hline                                                                           
Name & $t_{\rm mrg}/M$ & $f_{\rm mrg}$ [kHz]   \\
\hline
spALF2-1c & $3405$ & $1.62$   \\ 
irALF2    & $3536$ & $1.72$   \\  
coALF2    & $3630$ & $1.75$   \\  
spALF2+2c & $4607$ & $1.78$   \\  
spSLy-1c  & $3495$ & $1.47$   \\  
irSLy     & $3630$ & $1.98$   \\
coSly     & $3724$ & $2.00$   \\  
spSLy+2c  & $4262$ & $2.15$   \\  
\hline
\hline                                                                           
\end{tabular}                                                                                                                      
\end{table}  

\begin{figure}                                                                   
\begin{center}                                                                   
\includegraphics[width=0.99\columnwidth]{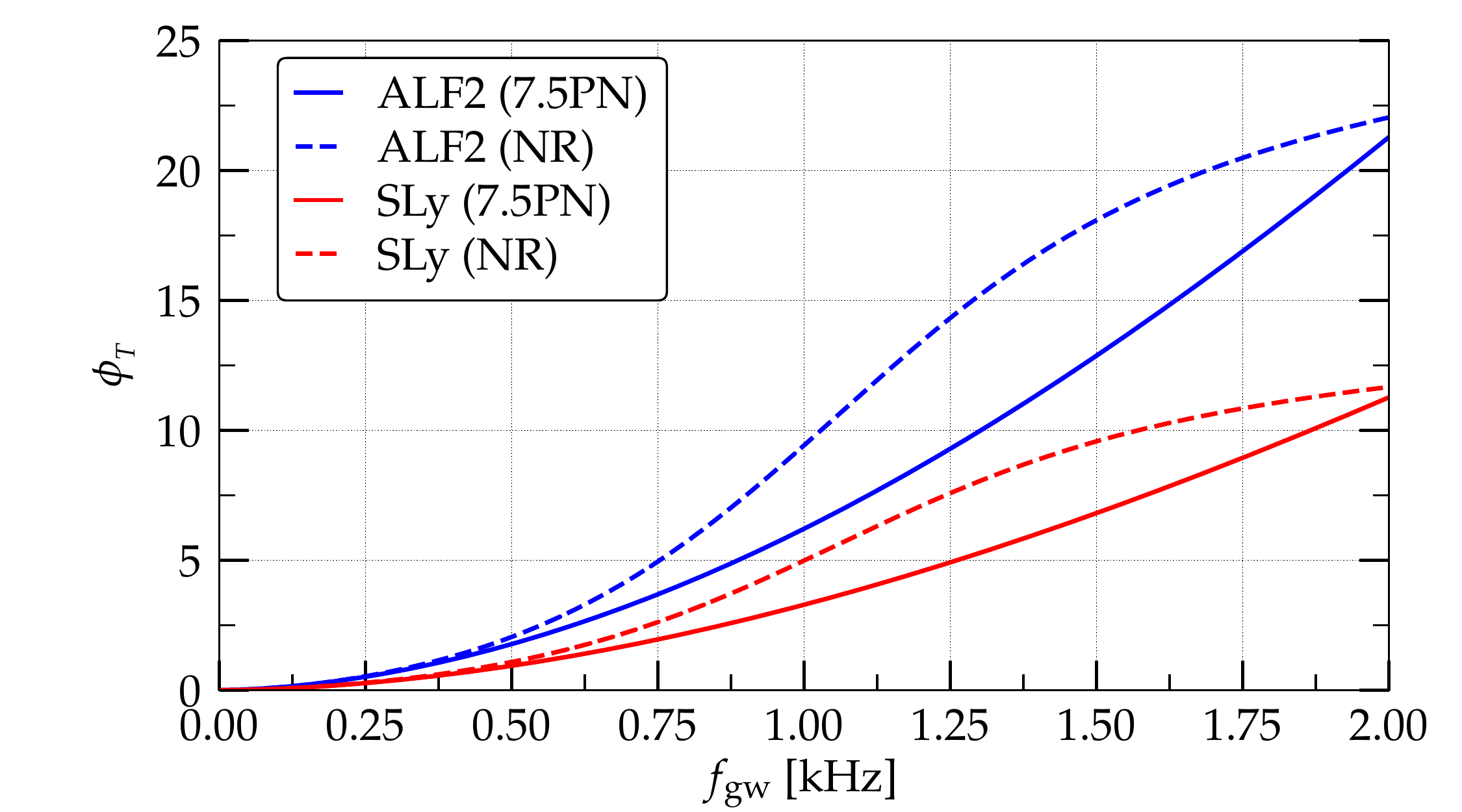}                       
\caption{The 7.5PN tidal part (solid lines) of the GW phase for 
the two EoSs used in this work, together with the tidal approximant of Dietrich 
et al. \cite{PhysRevD.96.121501} (dashed lines).}                                                           
\label{fig:PNphase}                                                               
\end{center}                                                                     
\end{figure}

To see this, note that tidal contributions enter the GW phase at the
5PN order and are partially known up to 7.5PN
\cite{PhysRevD.85.123007},
\begin{eqnarray}
\phi_{_T} & = & \sum_{i=1}^2 \kappa_i c_{_{\rm Newt}}^i x^{5/2}
(1+c_1^i x + c_{3/2}^i x^{3/2} + c_2^i x^2\nonumber   \\
         &    &\qquad\qquad\qquad\ \ + c_{5/2}^i x^{5/2})\ ,
\label{eq:phi7.5PN}
\end{eqnarray}
where $x=(M\pi f_{\rm gw})^{2/3}$ and the tidal deformability enters
through the coefficient $\kappa_1=3\Lambda_1 X_1^4 X_2$ (similarly for
$\kappa_2$). Here $X_i=M_i/M$, with $M_i$ the individual gravitational
masses, and all the coefficients $c^i$ are functions of $X_i$ (see
\cite{PhysRevD.85.123007}). Eq. \ref{eq:phi7.5PN} is plotted with
solid lines in Fig. \ref{fig:PNphase} for the two EoSs considered
here.

In addition to the PN formula, tidal effects can be described based on
numerical relativity simulations using the approximants derived in
Refs. \cite{PhysRevD.97.044044,PhysRevD.96.121501} either in the
frequency or in the time domain. The basic idea of these approximants
is to use binary black hole models in order to provide analytical
closed-form expressions correcting the GW phase to include tidal
effects. Here we use the approximant $\phi_{_T}^{_{\rm NR}}$
\cite{PhysRevD.96.121501,Dietrich:2019kaq} referred to as NRTidal,
which models the tidal effects in the time domain, Eq. \ref{eq:h1}.
In Fig. \ref{fig:PNphase} we plot $\phi_{_T}^{_{\rm NR}}$ (dashed
lines) with respect to the frequency for the ALF2 and SLy models used
in our simulation.  As shown in the plot, the NRTidal phase shift
between $f_{\rm gw}=0.4$ kHz and $f_{\rm gw}=1$ kHz is $\lesssim 8$
radian for ALF2 and $\lesssim 4$ radians for SLy. The aforementioned
NRTidal phase shift for SLy (ALF2) is comparable with the phase shift
due to spin for the SLy models spSLy-1c and coSly (ALF2 models
spALF2-1c and coALF2) as shown in the bottom row of
Fig. \ref{fig:phase}. For frequencies beyond the LIGO band, tidal
effects still prevail over the spin for those cases.

However, for our highest spinning models the picture is completely
different. At 1 KHz both the ALF2 and SLy EoSs develop a phase shift
due to spin approximately 4 times larger than the one coming from
tidal effects alone.  Even for larger frequencies the shift due to spin in
those cases will be larger than the corresponding one due to tidal
effects, despite the fact that the slopes of the curves of
Fig. \ref{fig:phase} are smaller than those of Fig. \ref{fig:PNphase}
in the 1-2 KHz regime.

Another interesting feature of Fig. \ref{fig:phase} is the fact that
for a given EoS the phase difference of a spinning model with respect
to the irrotational one \textit{does not} scale linearly with the
spin, a reminder of its nonlinear nature.  For example although the
antialigned and corotating ALF2 models have an absolute value of spin
which is approximately half of the spALF2+2c model, the phase
difference of the latter is approximately 8 times larger. For the SLy
EoS the antialigned and corotating models have an absolute value of
spin which is almost half of the spSLy+2c model, but the phase
difference of the latter is approximately 4 times larger. Also by
observing the antialigned and corotating ALF2, SLy models we can see
that they produce similar phase shifts with respect to the
irrotational case although their corresponding spins are $|\chi_{_{\rm
    ALF2}}|\approx 1.6 |\chi_{_{\rm SLy}}|$. In other words for
smaller spins softer EoSs produce the same phase shift as a stiffer
one with a higher spin. 

\begin{figure}                                                                   
\begin{center}                                                                   
\includegraphics[width=0.99\columnwidth]{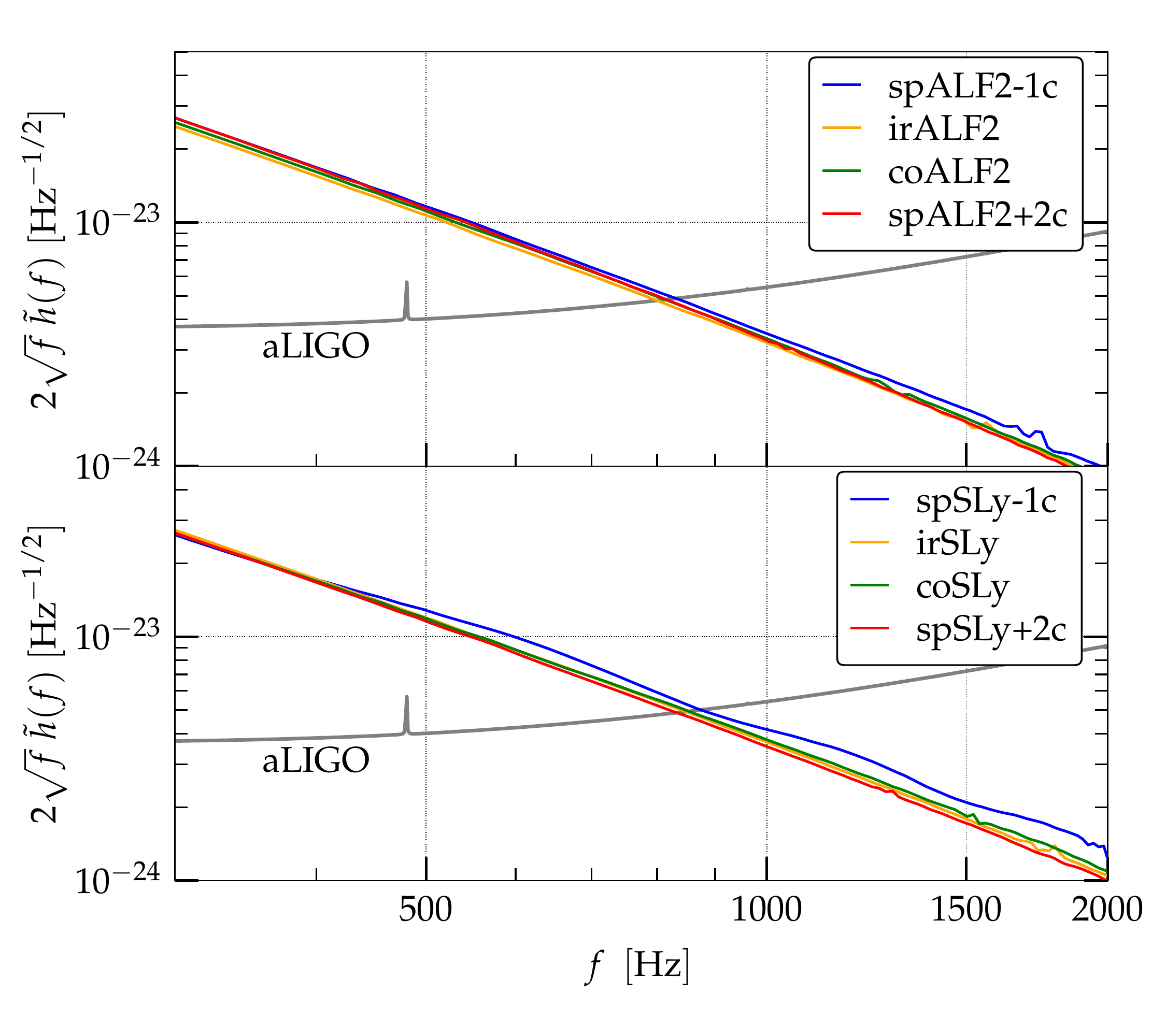}                       
\caption{GW spectra of the ALF2 and SLy numerical waveforms at
  $50$ Mpc, and aLIGO {\tt ZERO\_DET\_high\_P} noise curve (thick grey
  lines).}
\label{fig:fft}                                                               
\end{center}                                                                     
\end{figure}

The power spectral density of the models we simulated together with
the {\tt ZERO\_DET\_high\_P} aLIGO noise curve ($\sqrt{S_n(f)}$) are plotted in
Fig. \ref{fig:fft}. Spin effects are clearly not distinguishable on this plot.

\section{Discussion}

In this work we performed long-term inspiral simulations of
irrotational and highly spinning binary neutron stars using the
\illinois code in an effort to assess the importance of high spin. We
used two EoSs representing NSs of different compactions and three
different spins in order to compare the phase evolution with respect
to the irrotational case. Our spinning models range from binaries with
a spin $\sim 0.32$ aligned with the orbital angular momentum, to
antialigned binaries with a spin of $\sim -0.17$, all of them of equal
mass.  We employed high resolution with our finest grid spacing
$\Delta x=98$ m, motivated by the study of \cite{PhysRevD.96.084060}.
We find that our highest spinning binary exhibit a phase difference of
$\sim 40$ radians with respect to the irrotational one.  This shift
grows nonlinearly with the spin and depends on the EoS too. Our
findings indicate in full general relativity that \textit{the effect
  of moderate to high spin in the inspiral can be larger than the
  tidal effects alone, even when the rotation of the stars is far from
  their Keplerian limit}.  The dephasing due to spin is in accordance
with post-Newtonian analysis \cite{Harry2018}, and this work
underlines the importance of taking it into account for more reliable
GW data analysis.

Despite the fact that our calculations employ among the highest
resolutions adopted in numerical relativity simulations of inspiraling
binary neutron stars to-date, we find that our irrotational models
complete about 0.5-1 fewer orbits when compared to previous
studies. This would suggest a maximum phase error of about $4\pi\approx 12.6$
radians. To obtain a better handle on the phase error in our
calculations, and test whether a phase difference between a spin 0.32
and spin 0 binary can be as high as $\sim 40$ radians, we used the
{\tt IMRPhenomD} approximant~\cite{Khan:2015jqa} as implemented in
{\tt PyCBC}~\cite{Biwer:2018osg} to construct time domain binary black
hole waveforms. The results suggests that for a total mass of
2.72$M_\odot$, starting 400 Hz and ending at $1\ {\rm kHz}$, the phase
difference between an equal-mass, non-spinning binary black hole and a
binary black hole with dimensionless spin parameters $0.32$ is $\sim$
25 radians. This suggests that the phase difference of $\sim 40$
radians between our highest spin and irrotational ALF2 cases is likely
an overestimate, indicating that the phase error in our calculations
is possibly as high as $\sim 15$ radians for the ALF2 EoS. A similar
calculation for the spins we treat in the SLy EoS, shows a phase difference
in the binary black hole case of $\sim 15$ radians, suggesting an error in
our SLy phase difference calculations possibly as high as $5$ radians. Regardless,
the main result of our work is intact: the effect of spin in
  the inspiral of a binary neutron star system can be larger than the
  tidal effects and depends of the EoS, hence its inclusion in the GW
data analysis is important. While this result may sound obvious, we
point out that the spacetime outside a rotating NS is not Kerr, and
hence one cannot a-priori expect that spin effects in binary neutron
stars will be the same as those in binary black holes. Thus, our
calculations provide an explicit demonstration that spin effects can
be very important during the inspiral of a binary neutron star.

\acknowledgements
V.P. would like to thank KITP for hospitality, where part of this work
was completed. The authors would also like to thank D. Brown for help
with the PyCBC library.  This work has been supported in part by
National Science Foundation (NSF) Grant PHY-1662211, and NASA Grant
80NSSC17K0070 at the University of Illinois at Urbana-Champaign, as
well as by JSPS Grant-in-Aid for Scientific Research (C) 15K05085 and
18K03624 to the University of Ryukyus. KITP is supported in part by
the National Science Foundation under Grant No. NSF PHY-1748958. This
work made use of the Extreme Science and Engineering Discovery
Environment (XSEDE), which is supported by National Science Foundation
grant number TG-MCA99S008. This research is part of the Blue Waters
sustained-petascale computing project, which is supported by the
National Science Foundation (awards OCI-0725070 and ACI-1238993) and
the State of Illinois. Blue Waters is a joint effort of the University
of Illinois at Urbana-Champaign and its National Center for Supercomputing
Applications. Resources supporting this work were also provided by the NASA
High-End Computing (HEC) Program through the NASA Advanced Supercomputing
(NAS) Division at Ames Research Center.

\bibliography{references}

\end{document}

%% file: kigou.tex
\newcommand{\beq}{\begin{equation}} 
\newcommand{\eeq}{\end{equation}} 
\newcommand{\beqn}{\begin{eqnarray}} 
\newcommand{\eeqn}{\end{eqnarray}}

\newcommand{\zD}{{\raise1.0ex\hbox{${}^{\ \circ}$}}\!\!\!\!\!D}
\newcommand{\alone}{{\raise0.5ex\hbox{${}^{\ 1}$}}\!\!\!\!\alpha}

\newcommand{\nalam}{\mathrel{\raise0.9ex\hbox{$^\lambda$}\mkern-14mu
\lower0.0ex\hbox{$\nabla$}}}

\newcommand{\zeroD}{{\raise1.0ex\hbox{${}^{\ \circ}$}}\!\!\!\!\!D}

\newcommand{\zLap}{{\raise1.0ex\hbox{${}^{\ \circ}$}}\!\!\!\!\Delta}
\newcommand{\zna}{{\raise1.0ex\hbox{${}^{\ \circ}$}}\!\!\!\!\!\nabla}
\newcommand{\zS}{{\raise1.0ex\hbox{${}^{\ \circ}$}}\!\!\!\!\!S}

\newcommand{\carpet}{\textsc{\scalefont{1.1}carpet}\xspace}
\newcommand{\cocal}{\textsc{\scalefont{1.1}cocal}\xspace}

\newcommand{\sgrid}{\textsc{\scalefont{1.1}sgrid}\xspace}
\newcommand{\cactus}{\textsc{\scalefont{1.1}cactus}\xspace}
\newcommand{\bam}{\textsc{\scalefont{1.1}bam}\xspace}
\newcommand{\spec}{\textsc{\scalefont{1.1}s}p\textsc{\scalefont{1.1}ec}\xspace}

\newcommand{\whiskythc}{\textsc{\scalefont{1.1}WhiskyTHC}\xspace}

\newcommand{\illinois}{\textsc{\scalefont{1.1}Illinois grmhd}\xspace}

%% file: inspiral.bbl
\begin{thebibliography}{111}%
\makeatletter
\providecommand \@ifxundefined [1]{%
 \@ifx{#1\undefined}
}%
\providecommand \@ifnum [1]{%
 \ifnum #1\expandafter \@firstoftwo
 \else \expandafter \@secondoftwo
 \fi
}%
\providecommand \@ifx [1]{%
 \ifx #1\expandafter \@firstoftwo
 \else \expandafter \@secondoftwo
 \fi
}%
\providecommand \natexlab [1]{#1}%
\providecommand \enquote  [1]{``#1''}%
\providecommand \bibnamefont  [1]{#1}%
\providecommand \bibfnamefont [1]{#1}%
\providecommand \citenamefont [1]{#1}%
\providecommand \href@noop [0]{\@secondoftwo}%
\providecommand \href [0]{\begingroup \@sanitize@url \@href}%
\providecommand \@href[1]{\@@startlink{#1}\@@href}%
\providecommand \@@href[1]{\endgroup#1\@@endlink}%
\providecommand \@sanitize@url [0]{\catcode `\\12\catcode `\$12\catcode
  `\&12\catcode `\#12\catcode `\^12\catcode `\_12\catcode `\%12\relax}%
\providecommand \@@startlink[1]{}%
\providecommand \@@endlink[0]{}%
\providecommand \url  [0]{\begingroup\@sanitize@url \@url }%
\providecommand \@url [1]{\endgroup\@href {#1}{\urlprefix }}%
\providecommand \urlprefix  [0]{URL }%
\providecommand \Eprint [0]{\href }%
\providecommand \doibase [0]{http://dx.doi.org/}%
\providecommand \selectlanguage [0]{\@gobble}%
\providecommand \bibinfo  [0]{\@secondoftwo}%
\providecommand \bibfield  [0]{\@secondoftwo}%
\providecommand \translation [1]{[#1]}%
\providecommand \BibitemOpen [0]{}%
\providecommand \bibitemStop [0]{}%
\providecommand \bibitemNoStop [0]{.\EOS\space}%
\providecommand \EOS [0]{\spacefactor3000\relax}%
\providecommand \BibitemShut  [1]{\csname bibitem#1\endcsname}%
\let\auto@bib@innerbib\@empty
\bibitem [{\citenamefont {Aasi}\ \emph {et~al.}(2015)\citenamefont {Aasi} \emph
  {et~al.}}]{Aasi_2015}%
  \BibitemOpen
  \bibfield  {author} {\bibinfo {author} {\bibfnamefont {J.}~\bibnamefont
  {Aasi}} \emph {et~al.},\ }\href {\doibase 10.1088/0264-9381/32/11/115012}
  {\bibfield  {journal} {\bibinfo  {journal} {Classical and Quantum Gravity}\
  }\textbf {\bibinfo {volume} {32}},\ \bibinfo {pages} {115012} (\bibinfo
  {year} {2015})}\BibitemShut {NoStop}%
\bibitem [{\citenamefont {Acernese}\ \emph {et~al.}(2014)\citenamefont
  {Acernese} \emph {et~al.}}]{Acernese_2014}%
  \BibitemOpen
  \bibfield  {author} {\bibinfo {author} {\bibfnamefont {F.}~\bibnamefont
  {Acernese}} \emph {et~al.},\ }\href {\doibase 10.1088/0264-9381/32/2/024001}
  {\bibfield  {journal} {\bibinfo  {journal} {Classical and Quantum Gravity}\
  }\textbf {\bibinfo {volume} {32}},\ \bibinfo {pages} {024001} (\bibinfo
  {year} {2014})}\BibitemShut {NoStop}%
\bibitem [{\citenamefont {Abbott}\ \emph
  {et~al.}(2017{\natexlab{a}})\citenamefont {Abbott} \emph
  {et~al.}}]{TheLIGOScientific:2017qsa}%
  \BibitemOpen
  \bibfield  {author} {\bibinfo {author} {\bibfnamefont {B.~P.}\ \bibnamefont
  {Abbott}} \emph {et~al.} (\bibinfo {collaboration} {Virgo, LIGO
  Scientific}),\ }\href {\doibase 10.1103/PhysRevLett.119.161101} {\bibfield
  {journal} {\bibinfo  {journal} {Phys. Rev. Lett.}\ }\textbf {\bibinfo
  {volume} {119}},\ \bibinfo {pages} {161101} (\bibinfo {year}
  {2017}{\natexlab{a}})},\ \Eprint {http://arxiv.org/abs/1710.05832}
  {arXiv:1710.05832 [gr-qc]} \BibitemShut {NoStop}%
\bibitem [{GBM(2017)}]{GBM:2017lvd}%
  \BibitemOpen
  \href {\doibase 10.3847/2041-8213/aa91c9} {\enquote {\bibinfo {title}
  {{Multi-messenger Observations of a Binary Neutron Star Merger}},}\ }
  (\bibinfo {year} {2017}),\ \Eprint {http://arxiv.org/abs/1710.05833}
  {arXiv:1710.05833 [astro-ph.HE]} \BibitemShut {NoStop}%
\bibitem [{\citenamefont {Abbott}\ \emph
  {et~al.}(2017{\natexlab{b}})\citenamefont {Abbott} \emph
  {et~al.}}]{Monitor:2017mdv}%
  \BibitemOpen
  \bibfield  {author} {\bibinfo {author} {\bibfnamefont {B.~P.}\ \bibnamefont
  {Abbott}} \emph {et~al.} (\bibinfo {collaboration} {Virgo, Fermi-GBM,
  INTEGRAL, LIGO Scientific}),\ }\href {\doibase 10.3847/2041-8213/aa920c}
  {\bibfield  {journal} {\bibinfo  {journal} {Astrophys. J.}\ }\textbf
  {\bibinfo {volume} {848}},\ \bibinfo {pages} {L13} (\bibinfo {year}
  {2017}{\natexlab{b}})},\ \Eprint {http://arxiv.org/abs/1710.05834}
  {arXiv:1710.05834 [astro-ph.HE]} \BibitemShut {NoStop}%
\bibitem [{\citenamefont {Chornock}\ \emph {et~al.}(2017)\citenamefont
  {Chornock} \emph {et~al.}}]{Chornock:2017sdf}%
  \BibitemOpen
  \bibfield  {author} {\bibinfo {author} {\bibfnamefont {R.}~\bibnamefont
  {Chornock}} \emph {et~al.},\ }\href {\doibase 10.3847/2041-8213/aa905c}
  {\bibfield  {journal} {\bibinfo  {journal} {Astrophys. J.}\ }\textbf
  {\bibinfo {volume} {848}},\ \bibinfo {pages} {L19} (\bibinfo {year}
  {2017})},\ \Eprint {http://arxiv.org/abs/1710.05454} {arXiv:1710.05454
  [astro-ph.HE]} \BibitemShut {NoStop}%
\bibitem [{\citenamefont {{von Kienlin}}\ \emph {et~al.}(2017)\citenamefont
  {{von Kienlin}}, \citenamefont {{Meegan}},\ and\ \citenamefont
  {{Goldstein}}}]{2017GCN.21520....1V}%
  \BibitemOpen
  \bibfield  {author} {\bibinfo {author} {\bibfnamefont {A.}~\bibnamefont {{von
  Kienlin}}}, \bibinfo {author} {\bibfnamefont {C.}~\bibnamefont {{Meegan}}}, \
  and\ \bibinfo {author} {\bibfnamefont {A.}~\bibnamefont {{Goldstein}}},\
  }\href@noop {} {\bibfield  {journal} {\bibinfo  {journal} {GRB Coordinates
  Network, Circular Service, No.~21520, \#1 (2017)}\ }\textbf {\bibinfo
  {volume} {1520}} (\bibinfo {year} {2017})}\BibitemShut {NoStop}%
\bibitem [{\citenamefont {Savchenko}\ \emph {et~al.}(2017)\citenamefont
  {Savchenko} \emph {et~al.}}]{Savchenko:2017ffs}%
  \BibitemOpen
  \bibfield  {author} {\bibinfo {author} {\bibfnamefont {V.}~\bibnamefont
  {Savchenko}} \emph {et~al.},\ }\href {\doibase 10.3847/2041-8213/aa8f94}
  {\bibfield  {journal} {\bibinfo  {journal} {Astrophys. J.}\ }\textbf
  {\bibinfo {volume} {848}},\ \bibinfo {pages} {L15} (\bibinfo {year}
  {2017})},\ \Eprint {http://arxiv.org/abs/1710.05449} {arXiv:1710.05449
  [astro-ph.HE]} \BibitemShut {NoStop}%
\bibitem [{\citenamefont {Hinderer}\ \emph {et~al.}(2018)\citenamefont
  {Hinderer} \emph {et~al.}}]{Hinderer:2018pei}%
  \BibitemOpen
  \bibfield  {author} {\bibinfo {author} {\bibfnamefont {T.}~\bibnamefont
  {Hinderer}} \emph {et~al.},\ }\href@noop {} {\  (\bibinfo {year} {2018})},\
  \Eprint {http://arxiv.org/abs/1808.03836} {arXiv:1808.03836 [astro-ph.HE]}
  \BibitemShut {NoStop}%
\bibitem [{\citenamefont {Lattimer}\ and\ \citenamefont
  {Prakash}(2004)}]{Lattimer:2004pg}%
  \BibitemOpen
  \bibfield  {author} {\bibinfo {author} {\bibfnamefont {J.~M.}\ \bibnamefont
  {Lattimer}}\ and\ \bibinfo {author} {\bibfnamefont {M.}~\bibnamefont
  {Prakash}},\ }\href {\doibase 10.1126/science.1090720} {\bibfield  {journal}
  {\bibinfo  {journal} {Science}\ }\textbf {\bibinfo {volume} {304}},\ \bibinfo
  {pages} {536} (\bibinfo {year} {2004})},\ \Eprint
  {http://arxiv.org/abs/astro-ph/0405262} {arXiv:astro-ph/0405262 [astro-ph]}
  \BibitemShut {NoStop}%
\bibitem [{\citenamefont {Lattimer}(2012)}]{Lattimer:2012nd}%
  \BibitemOpen
  \bibfield  {author} {\bibinfo {author} {\bibfnamefont {J.~M.}\ \bibnamefont
  {Lattimer}},\ }\href {\doibase 10.1146/annurev-nucl-102711-095018} {\bibfield
   {journal} {\bibinfo  {journal} {Ann. Rev. Nucl. Part. Sci.}\ }\textbf
  {\bibinfo {volume} {62}},\ \bibinfo {pages} {485} (\bibinfo {year} {2012})},\
  \Eprint {http://arxiv.org/abs/1305.3510} {arXiv:1305.3510 [nucl-th]}
  \BibitemShut {NoStop}%
\bibitem [{\citenamefont {Özel}\ and\ \citenamefont
  {Freire}(2016)}]{Ozel:2016oaf}%
  \BibitemOpen
  \bibfield  {author} {\bibinfo {author} {\bibfnamefont {F.}~\bibnamefont
  {Özel}}\ and\ \bibinfo {author} {\bibfnamefont {P.}~\bibnamefont {Freire}},\
  }\href {\doibase 10.1146/annurev-astro-081915-023322} {\bibfield  {journal}
  {\bibinfo  {journal} {Ann. Rev. Astron. Astrophys.}\ }\textbf {\bibinfo
  {volume} {54}},\ \bibinfo {pages} {401} (\bibinfo {year} {2016})},\ \Eprint
  {http://arxiv.org/abs/1603.02698} {arXiv:1603.02698 [astro-ph.HE]}
  \BibitemShut {NoStop}%
\bibitem [{\citenamefont {Baiotti}\ and\ \citenamefont
  {Rezzolla}(2017)}]{Baiotti:2016qnr}%
  \BibitemOpen
  \bibfield  {author} {\bibinfo {author} {\bibfnamefont {L.}~\bibnamefont
  {Baiotti}}\ and\ \bibinfo {author} {\bibfnamefont {L.}~\bibnamefont
  {Rezzolla}},\ }\href {\doibase 10.1088/1361-6633/aa67bb} {\bibfield
  {journal} {\bibinfo  {journal} {Rept. Prog. Phys.}\ }\textbf {\bibinfo
  {volume} {80}},\ \bibinfo {pages} {096901} (\bibinfo {year} {2017})},\
  \Eprint {http://arxiv.org/abs/1607.03540} {arXiv:1607.03540 [gr-qc]}
  \BibitemShut {NoStop}%
\bibitem [{\citenamefont {Paschalidis}\ and\ \citenamefont
  {Stergioulas}(2017)}]{Paschalidis:2016vmz}%
  \BibitemOpen
  \bibfield  {author} {\bibinfo {author} {\bibfnamefont {V.}~\bibnamefont
  {Paschalidis}}\ and\ \bibinfo {author} {\bibfnamefont {N.}~\bibnamefont
  {Stergioulas}},\ }\href {\doibase 10.1007/s41114-017-0008-x} {\bibfield
  {journal} {\bibinfo  {journal} {Living Rev. Rel.}\ }\textbf {\bibinfo
  {volume} {20}},\ \bibinfo {pages} {7} (\bibinfo {year} {2017})},\ \Eprint
  {http://arxiv.org/abs/1612.03050} {arXiv:1612.03050 [astro-ph.HE]}
  \BibitemShut {NoStop}%
\bibitem [{\citenamefont {Cutler}\ and\ \citenamefont
  {Flanagan}(1994)}]{Cutler:1994ys}%
  \BibitemOpen
  \bibfield  {author} {\bibinfo {author} {\bibfnamefont {C.}~\bibnamefont
  {Cutler}}\ and\ \bibinfo {author} {\bibfnamefont {E.~E.}\ \bibnamefont
  {Flanagan}},\ }\href {\doibase 10.1103/PhysRevD.49.2658} {\bibfield
  {journal} {\bibinfo  {journal} {Phys. Rev.}\ }\textbf {\bibinfo {volume}
  {D49}},\ \bibinfo {pages} {2658} (\bibinfo {year} {1994})}\BibitemShut
  {NoStop}%
\bibitem [{\citenamefont {{Lai}}\ \emph {et~al.}(1994)\citenamefont {{Lai}},
  \citenamefont {{Rasio}},\ and\ \citenamefont
  {{Shapiro}}}]{1994ApJ...420..811L}%
  \BibitemOpen
  \bibfield  {author} {\bibinfo {author} {\bibfnamefont {D.}~\bibnamefont
  {{Lai}}}, \bibinfo {author} {\bibfnamefont {F.~A.}\ \bibnamefont {{Rasio}}},
  \ and\ \bibinfo {author} {\bibfnamefont {S.~L.}\ \bibnamefont {{Shapiro}}},\
  }\href {\doibase 10.1086/173606} {\bibfield  {journal} {\bibinfo  {journal}
  {\apj}\ }\textbf {\bibinfo {volume} {420}},\ \bibinfo {pages} {811} (\bibinfo
  {year} {1994})},\ \Eprint {http://arxiv.org/abs/astro-ph/9304027}
  {astro-ph/9304027} \BibitemShut {NoStop}%
\bibitem [{\citenamefont {Flanagan}\ and\ \citenamefont
  {Hinderer}(2008)}]{PhysRevD.77.021502}%
  \BibitemOpen
  \bibfield  {author} {\bibinfo {author} {\bibfnamefont {E.~E.}\ \bibnamefont
  {Flanagan}}\ and\ \bibinfo {author} {\bibfnamefont {T.}~\bibnamefont
  {Hinderer}},\ }\href {\doibase 10.1103/PhysRevD.77.021502} {\bibfield
  {journal} {\bibinfo  {journal} {Phys. Rev. D}\ }\textbf {\bibinfo {volume}
  {77}},\ \bibinfo {pages} {021502} (\bibinfo {year} {2008})}\BibitemShut
  {NoStop}%
\bibitem [{\citenamefont {Vines}\ \emph {et~al.}(2011)\citenamefont {Vines},
  \citenamefont {Flanagan},\ and\ \citenamefont
  {Hinderer}}]{PhysRevD.83.084051}%
  \BibitemOpen
  \bibfield  {author} {\bibinfo {author} {\bibfnamefont {J.}~\bibnamefont
  {Vines}}, \bibinfo {author} {\bibfnamefont {E.~E.}\ \bibnamefont {Flanagan}},
  \ and\ \bibinfo {author} {\bibfnamefont {T.}~\bibnamefont {Hinderer}},\
  }\href {\doibase 10.1103/PhysRevD.83.084051} {\bibfield  {journal} {\bibinfo
  {journal} {Phys. Rev. D}\ }\textbf {\bibinfo {volume} {83}},\ \bibinfo
  {pages} {084051} (\bibinfo {year} {2011})}\BibitemShut {NoStop}%
\bibitem [{\citenamefont {Hinderer}\ \emph {et~al.}(2010)\citenamefont
  {Hinderer}, \citenamefont {Lackey}, \citenamefont {Lang},\ and\ \citenamefont
  {Read}}]{Hinderer:2009ca}%
  \BibitemOpen
  \bibfield  {author} {\bibinfo {author} {\bibfnamefont {T.}~\bibnamefont
  {Hinderer}}, \bibinfo {author} {\bibfnamefont {B.~D.}\ \bibnamefont
  {Lackey}}, \bibinfo {author} {\bibfnamefont {R.~N.}\ \bibnamefont {Lang}}, \
  and\ \bibinfo {author} {\bibfnamefont {J.~S.}\ \bibnamefont {Read}},\ }\href
  {\doibase 10.1103/PhysRevD.81.123016} {\bibfield  {journal} {\bibinfo
  {journal} {Phys. Rev.}\ }\textbf {\bibinfo {volume} {D81}},\ \bibinfo {pages}
  {123016} (\bibinfo {year} {2010})}\BibitemShut {NoStop}%
\bibitem [{\citenamefont {Bini}\ \emph {et~al.}(2012)\citenamefont {Bini},
  \citenamefont {Damour},\ and\ \citenamefont {Faye}}]{PhysRevD.85.124034}%
  \BibitemOpen
  \bibfield  {author} {\bibinfo {author} {\bibfnamefont {D.}~\bibnamefont
  {Bini}}, \bibinfo {author} {\bibfnamefont {T.}~\bibnamefont {Damour}}, \ and\
  \bibinfo {author} {\bibfnamefont {G.}~\bibnamefont {Faye}},\ }\href {\doibase
  10.1103/PhysRevD.85.124034} {\bibfield  {journal} {\bibinfo  {journal} {Phys.
  Rev. D}\ }\textbf {\bibinfo {volume} {85}},\ \bibinfo {pages} {124034}
  (\bibinfo {year} {2012})}\BibitemShut {NoStop}%
\bibitem [{\citenamefont {Read}\ \emph
  {et~al.}(2013{\natexlab{a}})\citenamefont {Read}, \citenamefont {Baiotti},
  \citenamefont {Creighton}, \citenamefont {Friedman}, \citenamefont
  {Giacomazzo}, \citenamefont {Kyutoku}, \citenamefont {Markakis},
  \citenamefont {Rezzolla}, \citenamefont {Shibata},\ and\ \citenamefont
  {Taniguchi}}]{PhysRevD.88.044042}%
  \BibitemOpen
  \bibfield  {author} {\bibinfo {author} {\bibfnamefont {J.~S.}\ \bibnamefont
  {Read}}, \bibinfo {author} {\bibfnamefont {L.}~\bibnamefont {Baiotti}},
  \bibinfo {author} {\bibfnamefont {J.~D.~E.}\ \bibnamefont {Creighton}},
  \bibinfo {author} {\bibfnamefont {J.~L.}\ \bibnamefont {Friedman}}, \bibinfo
  {author} {\bibfnamefont {B.}~\bibnamefont {Giacomazzo}}, \bibinfo {author}
  {\bibfnamefont {K.}~\bibnamefont {Kyutoku}}, \bibinfo {author} {\bibfnamefont
  {C.}~\bibnamefont {Markakis}}, \bibinfo {author} {\bibfnamefont
  {L.}~\bibnamefont {Rezzolla}}, \bibinfo {author} {\bibfnamefont
  {M.}~\bibnamefont {Shibata}}, \ and\ \bibinfo {author} {\bibfnamefont
  {K.}~\bibnamefont {Taniguchi}},\ }\href {\doibase 10.1103/PhysRevD.88.044042}
  {\bibfield  {journal} {\bibinfo  {journal} {Phys. Rev. D}\ }\textbf {\bibinfo
  {volume} {88}},\ \bibinfo {pages} {044042} (\bibinfo {year}
  {2013}{\natexlab{a}})}\BibitemShut {NoStop}%
\bibitem [{\citenamefont {Lorimer}(2008)}]{lrr-2008-8}%
  \BibitemOpen
  \bibfield  {author} {\bibinfo {author} {\bibfnamefont {D.~R.}\ \bibnamefont
  {Lorimer}},\ }\href {\doibase 10.12942/lrr-2008-8} {\bibfield  {journal}
  {\bibinfo  {journal} {Living Reviews in Relativity}\ }\textbf {\bibinfo
  {volume} {11}} (\bibinfo {year} {2008}),\ 10.12942/lrr-2008-8}\BibitemShut
  {NoStop}%
\bibitem [{\citenamefont {Harry}\ and\ \citenamefont
  {Hinderer}(2018)}]{Harry2018}%
  \BibitemOpen
  \bibfield  {author} {\bibinfo {author} {\bibfnamefont {I.}~\bibnamefont
  {Harry}}\ and\ \bibinfo {author} {\bibfnamefont {T.}~\bibnamefont
  {Hinderer}},\ }\href {\doibase 10.1088/1361-6382/aac7e3} {\bibfield
  {journal} {\bibinfo  {journal} {Class. Quant. Grav.}\ }\textbf {\bibinfo
  {volume} {35}},\ \bibinfo {pages} {145010} (\bibinfo {year} {2018})},\
  \Eprint {http://arxiv.org/abs/1801.09972} {arXiv:1801.09972 [gr-qc]}
  \BibitemShut {NoStop}%
\bibitem [{\citenamefont {{Hessels}}\ \emph {et~al.}(2006)\citenamefont
  {{Hessels}}, \citenamefont {{Ransom}}, \citenamefont {{Stairs}},
  \citenamefont {{Freire}}, \citenamefont {{Kaspi}},\ and\ \citenamefont
  {{Camilo}}}]{2006Sci...311.1901H}%
  \BibitemOpen
  \bibfield  {author} {\bibinfo {author} {\bibfnamefont {J.~W.~T.}\
  \bibnamefont {{Hessels}}}, \bibinfo {author} {\bibfnamefont {S.~M.}\
  \bibnamefont {{Ransom}}}, \bibinfo {author} {\bibfnamefont {I.~H.}\
  \bibnamefont {{Stairs}}}, \bibinfo {author} {\bibfnamefont {P.~C.~C.}\
  \bibnamefont {{Freire}}}, \bibinfo {author} {\bibfnamefont {V.~M.}\
  \bibnamefont {{Kaspi}}}, \ and\ \bibinfo {author} {\bibfnamefont
  {F.}~\bibnamefont {{Camilo}}},\ }\href {\doibase 10.1126/science.1123430}
  {\bibfield  {journal} {\bibinfo  {journal} {Science}\ }\textbf {\bibinfo
  {volume} {311}},\ \bibinfo {pages} {1901} (\bibinfo {year} {2006})},\ \Eprint
  {http://arxiv.org/abs/astro-ph/0601337} {astro-ph/0601337} \BibitemShut
  {NoStop}%
\bibitem [{\citenamefont {Tauris}\ \emph {et~al.}(2017)\citenamefont {Tauris}
  \emph {et~al.}}]{Tauris:2017omb}%
  \BibitemOpen
  \bibfield  {author} {\bibinfo {author} {\bibfnamefont {T.~M.}\ \bibnamefont
  {Tauris}} \emph {et~al.},\ }\href {\doibase 10.3847/1538-4357/aa7e89}
  {\bibfield  {journal} {\bibinfo  {journal} {Astrophys. J.}\ }\textbf
  {\bibinfo {volume} {846}},\ \bibinfo {pages} {170} (\bibinfo {year}
  {2017})},\ \Eprint {http://arxiv.org/abs/1706.09438} {arXiv:1706.09438
  [astro-ph.HE]} \BibitemShut {NoStop}%
\bibitem [{\citenamefont {Zhu}\ \emph {et~al.}(2018)\citenamefont {Zhu},
  \citenamefont {Thrane}, \citenamefont {Os\l{}owski}, \citenamefont {Levin},\
  and\ \citenamefont {Lasky}}]{PhysRevD.98.043002}%
  \BibitemOpen
  \bibfield  {author} {\bibinfo {author} {\bibfnamefont {X.}~\bibnamefont
  {Zhu}}, \bibinfo {author} {\bibfnamefont {E.}~\bibnamefont {Thrane}},
  \bibinfo {author} {\bibfnamefont {S.}~\bibnamefont {Os\l{}owski}}, \bibinfo
  {author} {\bibfnamefont {Y.}~\bibnamefont {Levin}}, \ and\ \bibinfo {author}
  {\bibfnamefont {P.~D.}\ \bibnamefont {Lasky}},\ }\href {\doibase
  10.1103/PhysRevD.98.043002} {\bibfield  {journal} {\bibinfo  {journal} {Phys.
  Rev. D}\ }\textbf {\bibinfo {volume} {98}},\ \bibinfo {pages} {043002}
  (\bibinfo {year} {2018})}\BibitemShut {NoStop}%
\bibitem [{\citenamefont {Stovall}\ \emph {et~al.}(2018)\citenamefont {Stovall}
  \emph {et~al.}}]{Stovall:2018ouw}%
  \BibitemOpen
  \bibfield  {author} {\bibinfo {author} {\bibfnamefont {K.}~\bibnamefont
  {Stovall}} \emph {et~al.},\ }\href {\doibase 10.3847/2041-8213/aaad06}
  {\bibfield  {journal} {\bibinfo  {journal} {Astrophys. J.}\ }\textbf
  {\bibinfo {volume} {854}},\ \bibinfo {pages} {L22} (\bibinfo {year}
  {2018})},\ \Eprint {http://arxiv.org/abs/1802.01707} {arXiv:1802.01707
  [astro-ph.HE]} \BibitemShut {NoStop}%
\bibitem [{\citenamefont {Cameron}\ \emph {et~al.}(2018)\citenamefont {Cameron}
  \emph {et~al.}}]{Cameron:2017ody}%
  \BibitemOpen
  \bibfield  {author} {\bibinfo {author} {\bibfnamefont {A.~D.}\ \bibnamefont
  {Cameron}} \emph {et~al.},\ }\href {\doibase 10.1093/mnrasl/sly003}
  {\bibfield  {journal} {\bibinfo  {journal} {Mon. Not. Roy. Astron. Soc.}\
  }\textbf {\bibinfo {volume} {475}},\ \bibinfo {pages} {L57} (\bibinfo {year}
  {2018})},\ \Eprint {http://arxiv.org/abs/1711.07697} {arXiv:1711.07697
  [astro-ph.HE]} \BibitemShut {NoStop}%
\bibitem [{\citenamefont {Kramer}\ \emph {et~al.}(2006)\citenamefont {Kramer}
  \emph {et~al.}}]{Kramer:2006nb}%
  \BibitemOpen
  \bibfield  {author} {\bibinfo {author} {\bibfnamefont {M.}~\bibnamefont
  {Kramer}} \emph {et~al.},\ }\href {\doibase 10.1126/science.1132305}
  {\bibfield  {journal} {\bibinfo  {journal} {Science}\ }\textbf {\bibinfo
  {volume} {314}},\ \bibinfo {pages} {97} (\bibinfo {year} {2006})},\ \Eprint
  {http://arxiv.org/abs/astro-ph/0609417} {arXiv:astro-ph/0609417 [astro-ph]}
  \BibitemShut {NoStop}%
\bibitem [{\citenamefont {Tichy}(2011)}]{Tichy:2011gw}%
  \BibitemOpen
  \bibfield  {author} {\bibinfo {author} {\bibfnamefont {W.}~\bibnamefont
  {Tichy}},\ }\href {\doibase 10.1103/PhysRevD.84.024041} {\bibfield  {journal}
  {\bibinfo  {journal} {Phys. Rev.}\ }\textbf {\bibinfo {volume} {D84}},\
  \bibinfo {pages} {024041} (\bibinfo {year} {2011})},\ \Eprint
  {http://arxiv.org/abs/1107.1440} {arXiv:1107.1440 [gr-qc]} \BibitemShut
  {NoStop}%
\bibitem [{\citenamefont {{Tichy}}(2012)}]{2012PhRvD..86f4024T}%
  \BibitemOpen
  \bibfield  {author} {\bibinfo {author} {\bibfnamefont {W.}~\bibnamefont
  {{Tichy}}},\ }\href {\doibase 10.1103/PhysRevD.86.064024} {\bibfield
  {journal} {\bibinfo  {journal} {\prd}\ }\textbf {\bibinfo {volume} {86}},\
  \bibinfo {eid} {064024} (\bibinfo {year} {2012})},\ \Eprint
  {http://arxiv.org/abs/1209.5336} {arXiv:1209.5336} \BibitemShut {NoStop}%
\bibitem [{\citenamefont {Bernuzzi}\ \emph
  {et~al.}(2014{\natexlab{a}})\citenamefont {Bernuzzi}, \citenamefont
  {Dietrich}, \citenamefont {Tichy},\ and\ \citenamefont
  {Brügmann}}]{Bernuzzi:2013rza}%
  \BibitemOpen
  \bibfield  {author} {\bibinfo {author} {\bibfnamefont {S.}~\bibnamefont
  {Bernuzzi}}, \bibinfo {author} {\bibfnamefont {T.}~\bibnamefont {Dietrich}},
  \bibinfo {author} {\bibfnamefont {W.}~\bibnamefont {Tichy}}, \ and\ \bibinfo
  {author} {\bibfnamefont {B.}~\bibnamefont {Brügmann}},\ }\href {\doibase
  10.1103/PhysRevD.89.104021} {\bibfield  {journal} {\bibinfo  {journal} {Phys.
  Rev.}\ }\textbf {\bibinfo {volume} {D89}},\ \bibinfo {pages} {104021}
  (\bibinfo {year} {2014}{\natexlab{a}})},\ \Eprint
  {http://arxiv.org/abs/1311.4443} {arXiv:1311.4443 [gr-qc]} \BibitemShut
  {NoStop}%
\bibitem [{\citenamefont {Tacik}\ \emph
  {et~al.}(2015{\natexlab{a}})\citenamefont {Tacik} \emph
  {et~al.}}]{PhysRevD.92.124012}%
  \BibitemOpen
  \bibfield  {author} {\bibinfo {author} {\bibfnamefont {N.}~\bibnamefont
  {Tacik}} \emph {et~al.},\ }\href {\doibase 10.1103/PhysRevD.92.124012}
  {\bibfield  {journal} {\bibinfo  {journal} {Phys. Rev. D}\ }\textbf {\bibinfo
  {volume} {92}},\ \bibinfo {pages} {124012} (\bibinfo {year}
  {2015}{\natexlab{a}})}\BibitemShut {NoStop}%
\bibitem [{\citenamefont {Kiuchi}\ \emph {et~al.}(2017)\citenamefont {Kiuchi},
  \citenamefont {Kawaguchi}, \citenamefont {Kyutoku}, \citenamefont
  {Sekiguchi}, \citenamefont {Shibata},\ and\ \citenamefont
  {Taniguchi}}]{PhysRevD.96.084060}%
  \BibitemOpen
  \bibfield  {author} {\bibinfo {author} {\bibfnamefont {K.}~\bibnamefont
  {Kiuchi}}, \bibinfo {author} {\bibfnamefont {K.}~\bibnamefont {Kawaguchi}},
  \bibinfo {author} {\bibfnamefont {K.}~\bibnamefont {Kyutoku}}, \bibinfo
  {author} {\bibfnamefont {Y.}~\bibnamefont {Sekiguchi}}, \bibinfo {author}
  {\bibfnamefont {M.}~\bibnamefont {Shibata}}, \ and\ \bibinfo {author}
  {\bibfnamefont {K.}~\bibnamefont {Taniguchi}},\ }\href {\doibase
  10.1103/PhysRevD.96.084060} {\bibfield  {journal} {\bibinfo  {journal} {Phys.
  Rev. D}\ }\textbf {\bibinfo {volume} {96}},\ \bibinfo {pages} {084060}
  (\bibinfo {year} {2017})}\BibitemShut {NoStop}%
\bibitem [{\citenamefont {Baiotti}\ \emph {et~al.}(2008)\citenamefont
  {Baiotti}, \citenamefont {Giacomazzo},\ and\ \citenamefont
  {Rezzolla}}]{Baiotti:2008ra}%
  \BibitemOpen
  \bibfield  {author} {\bibinfo {author} {\bibfnamefont {L.}~\bibnamefont
  {Baiotti}}, \bibinfo {author} {\bibfnamefont {B.}~\bibnamefont {Giacomazzo}},
  \ and\ \bibinfo {author} {\bibfnamefont {L.}~\bibnamefont {Rezzolla}},\
  }\href {\doibase 10.1103/PhysRevD.78.084033} {\bibfield  {journal} {\bibinfo
  {journal} {Phys. Rev.}\ }\textbf {\bibinfo {volume} {D78}},\ \bibinfo {pages}
  {084033} (\bibinfo {year} {2008})}\BibitemShut {NoStop}%
\bibitem [{\citenamefont {Baiotti}\ \emph {et~al.}(2010)\citenamefont
  {Baiotti}, \citenamefont {Damour}, \citenamefont {Giacomazzo}, \citenamefont
  {Nagar},\ and\ \citenamefont {Rezzolla}}]{Baiotti:2010xh}%
  \BibitemOpen
  \bibfield  {author} {\bibinfo {author} {\bibfnamefont {L.}~\bibnamefont
  {Baiotti}}, \bibinfo {author} {\bibfnamefont {T.}~\bibnamefont {Damour}},
  \bibinfo {author} {\bibfnamefont {B.}~\bibnamefont {Giacomazzo}}, \bibinfo
  {author} {\bibfnamefont {A.}~\bibnamefont {Nagar}}, \ and\ \bibinfo {author}
  {\bibfnamefont {L.}~\bibnamefont {Rezzolla}},\ }\href {\doibase
  10.1103/PhysRevLett.105.261101} {\bibfield  {journal} {\bibinfo  {journal}
  {Phys. Rev. Lett.}\ }\textbf {\bibinfo {volume} {105}},\ \bibinfo {pages}
  {261101} (\bibinfo {year} {2010})},\ \Eprint {http://arxiv.org/abs/1009.0521}
  {arXiv:1009.0521 [gr-qc]} \BibitemShut {NoStop}%
\bibitem [{\citenamefont {Hotokezaka}\ \emph {et~al.}(2013)\citenamefont
  {Hotokezaka}, \citenamefont {Kyutoku},\ and\ \citenamefont
  {Shibata}}]{PhysRevD.87.044001}%
  \BibitemOpen
  \bibfield  {author} {\bibinfo {author} {\bibfnamefont {K.}~\bibnamefont
  {Hotokezaka}}, \bibinfo {author} {\bibfnamefont {K.}~\bibnamefont {Kyutoku}},
  \ and\ \bibinfo {author} {\bibfnamefont {M.}~\bibnamefont {Shibata}},\ }\href
  {\doibase 10.1103/PhysRevD.87.044001} {\bibfield  {journal} {\bibinfo
  {journal} {Phys. Rev. D}\ }\textbf {\bibinfo {volume} {87}},\ \bibinfo
  {pages} {044001} (\bibinfo {year} {2013})}\BibitemShut {NoStop}%
\bibitem [{\citenamefont {Hotokezaka}\ \emph {et~al.}(2015)\citenamefont
  {Hotokezaka}, \citenamefont {Kyutoku}, \citenamefont {Okawa},\ and\
  \citenamefont {Shibata}}]{PhysRevD.91.064060}%
  \BibitemOpen
  \bibfield  {author} {\bibinfo {author} {\bibfnamefont {K.}~\bibnamefont
  {Hotokezaka}}, \bibinfo {author} {\bibfnamefont {K.}~\bibnamefont {Kyutoku}},
  \bibinfo {author} {\bibfnamefont {H.}~\bibnamefont {Okawa}}, \ and\ \bibinfo
  {author} {\bibfnamefont {M.}~\bibnamefont {Shibata}},\ }\href {\doibase
  10.1103/PhysRevD.91.064060} {\bibfield  {journal} {\bibinfo  {journal} {Phys.
  Rev. D}\ }\textbf {\bibinfo {volume} {91}},\ \bibinfo {pages} {064060}
  (\bibinfo {year} {2015})}\BibitemShut {NoStop}%
\bibitem [{\citenamefont {Hotokezaka}\ \emph {et~al.}(2016)\citenamefont
  {Hotokezaka}, \citenamefont {Kyutoku}, \citenamefont {Sekiguchi},\ and\
  \citenamefont {Shibata}}]{PhysRevD.93.064082}%
  \BibitemOpen
  \bibfield  {author} {\bibinfo {author} {\bibfnamefont {K.}~\bibnamefont
  {Hotokezaka}}, \bibinfo {author} {\bibfnamefont {K.}~\bibnamefont {Kyutoku}},
  \bibinfo {author} {\bibfnamefont {Y.-i.}\ \bibnamefont {Sekiguchi}}, \ and\
  \bibinfo {author} {\bibfnamefont {M.}~\bibnamefont {Shibata}},\ }\href
  {\doibase 10.1103/PhysRevD.93.064082} {\bibfield  {journal} {\bibinfo
  {journal} {Phys. Rev. D}\ }\textbf {\bibinfo {volume} {93}},\ \bibinfo
  {pages} {064082} (\bibinfo {year} {2016})}\BibitemShut {NoStop}%
\bibitem [{\citenamefont {Kawaguchi}\ \emph {et~al.}(2018)\citenamefont
  {Kawaguchi}, \citenamefont {Kiuchi}, \citenamefont {Kyutoku}, \citenamefont
  {Sekiguchi}, \citenamefont {Shibata},\ and\ \citenamefont
  {Taniguchi}}]{PhysRevD.97.044044}%
  \BibitemOpen
  \bibfield  {author} {\bibinfo {author} {\bibfnamefont {K.}~\bibnamefont
  {Kawaguchi}}, \bibinfo {author} {\bibfnamefont {K.}~\bibnamefont {Kiuchi}},
  \bibinfo {author} {\bibfnamefont {K.}~\bibnamefont {Kyutoku}}, \bibinfo
  {author} {\bibfnamefont {Y.}~\bibnamefont {Sekiguchi}}, \bibinfo {author}
  {\bibfnamefont {M.}~\bibnamefont {Shibata}}, \ and\ \bibinfo {author}
  {\bibfnamefont {K.}~\bibnamefont {Taniguchi}},\ }\href {\doibase
  10.1103/PhysRevD.97.044044} {\bibfield  {journal} {\bibinfo  {journal} {Phys.
  Rev. D}\ }\textbf {\bibinfo {volume} {97}},\ \bibinfo {pages} {044044}
  (\bibinfo {year} {2018})}\BibitemShut {NoStop}%
\bibitem [{\citenamefont {Haas}\ \emph {et~al.}(2016)\citenamefont {Haas},
  \citenamefont {Ott}, \citenamefont {Szilagyi}, \citenamefont {Kaplan},
  \citenamefont {Lippuner}, \citenamefont {Scheel}, \citenamefont {Barkett},
  \citenamefont {Muhlberger}, \citenamefont {Dietrich}, \citenamefont {Duez},
  \citenamefont {Foucart}, \citenamefont {Pfeiffer}, \citenamefont {Kidder},\
  and\ \citenamefont {Teukolsky}}]{PhysRevD.93.124062}%
  \BibitemOpen
  \bibfield  {author} {\bibinfo {author} {\bibfnamefont {R.}~\bibnamefont
  {Haas}}, \bibinfo {author} {\bibfnamefont {C.~D.}\ \bibnamefont {Ott}},
  \bibinfo {author} {\bibfnamefont {B.}~\bibnamefont {Szilagyi}}, \bibinfo
  {author} {\bibfnamefont {J.~D.}\ \bibnamefont {Kaplan}}, \bibinfo {author}
  {\bibfnamefont {J.}~\bibnamefont {Lippuner}}, \bibinfo {author}
  {\bibfnamefont {M.~A.}\ \bibnamefont {Scheel}}, \bibinfo {author}
  {\bibfnamefont {K.}~\bibnamefont {Barkett}}, \bibinfo {author} {\bibfnamefont
  {C.~D.}\ \bibnamefont {Muhlberger}}, \bibinfo {author} {\bibfnamefont
  {T.}~\bibnamefont {Dietrich}}, \bibinfo {author} {\bibfnamefont {M.~D.}\
  \bibnamefont {Duez}}, \bibinfo {author} {\bibfnamefont {F.}~\bibnamefont
  {Foucart}}, \bibinfo {author} {\bibfnamefont {H.~P.}\ \bibnamefont
  {Pfeiffer}}, \bibinfo {author} {\bibfnamefont {L.~E.}\ \bibnamefont
  {Kidder}}, \ and\ \bibinfo {author} {\bibfnamefont {S.~A.}\ \bibnamefont
  {Teukolsky}},\ }\href {\doibase 10.1103/PhysRevD.93.124062} {\bibfield
  {journal} {\bibinfo  {journal} {Phys. Rev. D}\ }\textbf {\bibinfo {volume}
  {93}},\ \bibinfo {pages} {124062} (\bibinfo {year} {2016})}\BibitemShut
  {NoStop}%
\bibitem [{\citenamefont {Dietrich}\ \emph
  {et~al.}(2017{\natexlab{a}})\citenamefont {Dietrich}, \citenamefont
  {Bernuzzi}, \citenamefont {Ujevic},\ and\ \citenamefont
  {Tichy}}]{PhysRevD.95.044045}%
  \BibitemOpen
  \bibfield  {author} {\bibinfo {author} {\bibfnamefont {T.}~\bibnamefont
  {Dietrich}}, \bibinfo {author} {\bibfnamefont {S.}~\bibnamefont {Bernuzzi}},
  \bibinfo {author} {\bibfnamefont {M.}~\bibnamefont {Ujevic}}, \ and\ \bibinfo
  {author} {\bibfnamefont {W.}~\bibnamefont {Tichy}},\ }\href {\doibase
  10.1103/PhysRevD.95.044045} {\bibfield  {journal} {\bibinfo  {journal} {Phys.
  Rev. D}\ }\textbf {\bibinfo {volume} {95}},\ \bibinfo {pages} {044045}
  (\bibinfo {year} {2017}{\natexlab{a}})}\BibitemShut {NoStop}%
\bibitem [{\citenamefont {Dietrich}\ \emph
  {et~al.}(2018{\natexlab{a}})\citenamefont {Dietrich}, \citenamefont
  {Bernuzzi}, \citenamefont {Br\"ugmann}, \citenamefont {Ujevic},\ and\
  \citenamefont {Tichy}}]{PhysRevD.97.064002}%
  \BibitemOpen
  \bibfield  {author} {\bibinfo {author} {\bibfnamefont {T.}~\bibnamefont
  {Dietrich}}, \bibinfo {author} {\bibfnamefont {S.}~\bibnamefont {Bernuzzi}},
  \bibinfo {author} {\bibfnamefont {B.}~\bibnamefont {Br\"ugmann}}, \bibinfo
  {author} {\bibfnamefont {M.}~\bibnamefont {Ujevic}}, \ and\ \bibinfo {author}
  {\bibfnamefont {W.}~\bibnamefont {Tichy}},\ }\href {\doibase
  10.1103/PhysRevD.97.064002} {\bibfield  {journal} {\bibinfo  {journal} {Phys.
  Rev. D}\ }\textbf {\bibinfo {volume} {97}},\ \bibinfo {pages} {064002}
  (\bibinfo {year} {2018}{\natexlab{a}})}\BibitemShut {NoStop}%
\bibitem [{\citenamefont {Dietrich}\ \emph
  {et~al.}(2018{\natexlab{b}})\citenamefont {Dietrich}, \citenamefont
  {Bernuzzi}, \citenamefont {Bruegmann},\ and\ \citenamefont
  {Tichy}}]{Dietrich:2018upm}%
  \BibitemOpen
  \bibfield  {author} {\bibinfo {author} {\bibfnamefont {T.}~\bibnamefont
  {Dietrich}}, \bibinfo {author} {\bibfnamefont {S.}~\bibnamefont {Bernuzzi}},
  \bibinfo {author} {\bibfnamefont {B.}~\bibnamefont {Bruegmann}}, \ and\
  \bibinfo {author} {\bibfnamefont {W.}~\bibnamefont {Tichy}},\ }in\ \href
  {\doibase 10.1109/PDP2018.2018.00113} {\emph {\bibinfo {booktitle}
  {{Proceedings, 26th Euromicro International Conference on Parallel,
  Distributed and Network-based Processing (PDP 2018): Cambridge, UK, March
  21-23, 2018}}}}\ (\bibinfo {year} {2018})\ pp.\ \bibinfo {pages} {682--689},\
  \Eprint {http://arxiv.org/abs/1803.07965} {arXiv:1803.07965 [gr-qc]}
  \BibitemShut {NoStop}%
\bibitem [{\citenamefont {Dietrich}\ \emph
  {et~al.}(2017{\natexlab{b}})\citenamefont {Dietrich}, \citenamefont
  {Bernuzzi},\ and\ \citenamefont {Tichy}}]{PhysRevD.96.121501}%
  \BibitemOpen
  \bibfield  {author} {\bibinfo {author} {\bibfnamefont {T.}~\bibnamefont
  {Dietrich}}, \bibinfo {author} {\bibfnamefont {S.}~\bibnamefont {Bernuzzi}},
  \ and\ \bibinfo {author} {\bibfnamefont {W.}~\bibnamefont {Tichy}},\ }\href
  {\doibase 10.1103/PhysRevD.96.121501} {\bibfield  {journal} {\bibinfo
  {journal} {Phys. Rev. D}\ }\textbf {\bibinfo {volume} {96}},\ \bibinfo
  {pages} {121501} (\bibinfo {year} {2017}{\natexlab{b}})}\BibitemShut
  {NoStop}%
\bibitem [{\citenamefont {{Kastaun}}\ \emph {et~al.}(2013)\citenamefont
  {{Kastaun}}, \citenamefont {{Galeazzi}}, \citenamefont {{Alic}},
  \citenamefont {{Rezzolla}},\ and\ \citenamefont {{Font}}}]{Kastaun2013}%
  \BibitemOpen
  \bibfield  {author} {\bibinfo {author} {\bibfnamefont {W.}~\bibnamefont
  {{Kastaun}}}, \bibinfo {author} {\bibfnamefont {F.}~\bibnamefont
  {{Galeazzi}}}, \bibinfo {author} {\bibfnamefont {D.}~\bibnamefont {{Alic}}},
  \bibinfo {author} {\bibfnamefont {L.}~\bibnamefont {{Rezzolla}}}, \ and\
  \bibinfo {author} {\bibfnamefont {J.~A.}\ \bibnamefont {{Font}}},\ }\href
  {\doibase 10.1103/PhysRevD.88.021501} {\bibfield  {journal} {\bibinfo
  {journal} {Phys. Rev. D}\ }\textbf {\bibinfo {volume} {88}},\ \bibinfo {eid}
  {021501} (\bibinfo {year} {2013})},\ \Eprint {http://arxiv.org/abs/1301.7348}
  {arXiv:1301.7348 [gr-qc]} \BibitemShut {NoStop}%
\bibitem [{\citenamefont {{Kastaun}}\ and\ \citenamefont
  {{Galeazzi}}(2015)}]{Kastaun2015}%
  \BibitemOpen
  \bibfield  {author} {\bibinfo {author} {\bibfnamefont {W.}~\bibnamefont
  {{Kastaun}}}\ and\ \bibinfo {author} {\bibfnamefont {F.}~\bibnamefont
  {{Galeazzi}}},\ }\href {\doibase 10.1103/PhysRevD.91.064027} {\bibfield
  {journal} {\bibinfo  {journal} {\prd}\ }\textbf {\bibinfo {volume} {91}},\
  \bibinfo {eid} {064027} (\bibinfo {year} {2015})},\ \Eprint
  {http://arxiv.org/abs/1411.7975} {arXiv:1411.7975 [gr-qc]} \BibitemShut
  {NoStop}%
\bibitem [{\citenamefont {Tacik}\ \emph
  {et~al.}(2015{\natexlab{b}})\citenamefont {Tacik} \emph
  {et~al.}}]{Tacik:2015tja}%
  \BibitemOpen
  \bibfield  {author} {\bibinfo {author} {\bibfnamefont {N.}~\bibnamefont
  {Tacik}} \emph {et~al.},\ }\href {\doibase 10.1103/PhysRevD.94.049903,
  10.1103/PhysRevD.92.124012} {\bibfield  {journal} {\bibinfo  {journal} {Phys.
  Rev.}\ }\textbf {\bibinfo {volume} {D92}},\ \bibinfo {pages} {124012}
  (\bibinfo {year} {2015}{\natexlab{b}})},\ \bibinfo {note} {[Erratum: Phys.
  Rev.D94,no.4,049903(2016)]},\ \Eprint {http://arxiv.org/abs/1508.06986}
  {arXiv:1508.06986 [gr-qc]} \BibitemShut {NoStop}%
\bibitem [{\citenamefont {Bauswein}\ \emph {et~al.}(2016)\citenamefont
  {Bauswein}, \citenamefont {Stergioulas},\ and\ \citenamefont
  {Janka}}]{bauswein2015exploring}%
  \BibitemOpen
  \bibfield  {author} {\bibinfo {author} {\bibfnamefont {A.}~\bibnamefont
  {Bauswein}}, \bibinfo {author} {\bibfnamefont {N.}~\bibnamefont
  {Stergioulas}}, \ and\ \bibinfo {author} {\bibfnamefont {H.-T.}\ \bibnamefont
  {Janka}},\ }\href {\doibase 10.1140/epja/i2016-16056-7} {\bibfield  {journal}
  {\bibinfo  {journal} {Eur. Phys. J.}\ }\textbf {\bibinfo {volume} {A52}},\
  \bibinfo {pages} {56} (\bibinfo {year} {2016})},\ \Eprint
  {http://arxiv.org/abs/1508.05493} {arXiv:1508.05493 [astro-ph.HE]}
  \BibitemShut {NoStop}%
\bibitem [{\citenamefont {Dietrich}\ \emph {et~al.}(2015)\citenamefont
  {Dietrich}, \citenamefont {Moldenhauer}, \citenamefont {Johnson-McDaniel},
  \citenamefont {Bernuzzi}, \citenamefont {Markakis}, \citenamefont
  {Brügmann},\ and\ \citenamefont {Tichy}}]{Dietrich:2015pxa}%
  \BibitemOpen
  \bibfield  {author} {\bibinfo {author} {\bibfnamefont {T.}~\bibnamefont
  {Dietrich}}, \bibinfo {author} {\bibfnamefont {N.}~\bibnamefont
  {Moldenhauer}}, \bibinfo {author} {\bibfnamefont {N.~K.}\ \bibnamefont
  {Johnson-McDaniel}}, \bibinfo {author} {\bibfnamefont {S.}~\bibnamefont
  {Bernuzzi}}, \bibinfo {author} {\bibfnamefont {C.~M.}\ \bibnamefont
  {Markakis}}, \bibinfo {author} {\bibfnamefont {B.}~\bibnamefont {Brügmann}},
  \ and\ \bibinfo {author} {\bibfnamefont {W.}~\bibnamefont {Tichy}},\ }\href
  {\doibase 10.1103/PhysRevD.92.124007} {\bibfield  {journal} {\bibinfo
  {journal} {Phys. Rev.}\ }\textbf {\bibinfo {volume} {D92}},\ \bibinfo {pages}
  {124007} (\bibinfo {year} {2015})},\ \Eprint
  {http://arxiv.org/abs/1507.07100} {arXiv:1507.07100 [gr-qc]} \BibitemShut
  {NoStop}%
\bibitem [{\citenamefont {{East}}\ \emph {et~al.}(2015)\citenamefont {{East}},
  \citenamefont {{Paschalidis}},\ and\ \citenamefont {{Pretorius}}}]{EPP2015}%
  \BibitemOpen
  \bibfield  {author} {\bibinfo {author} {\bibfnamefont {W.~E.}\ \bibnamefont
  {{East}}}, \bibinfo {author} {\bibfnamefont {V.}~\bibnamefont
  {{Paschalidis}}}, \ and\ \bibinfo {author} {\bibfnamefont {F.}~\bibnamefont
  {{Pretorius}}},\ }\href {\doibase 10.1088/2041-8205/807/1/L3} {\bibfield
  {journal} {\bibinfo  {journal} {The Astrophysical Journal Letters}\ }\textbf
  {\bibinfo {volume} {807}},\ \bibinfo {eid} {L3} (\bibinfo {year} {2015})},\
  \Eprint {http://arxiv.org/abs/1503.07171} {arXiv:1503.07171 [astro-ph.HE]}
  \BibitemShut {NoStop}%
\bibitem [{\citenamefont {{Paschalidis}}\ \emph
  {et~al.}(2015{\natexlab{a}})\citenamefont {{Paschalidis}}, \citenamefont
  {{East}}, \citenamefont {{Pretorius}},\ and\ \citenamefont
  {{Shapiro}}}]{PEPS2015}%
  \BibitemOpen
  \bibfield  {author} {\bibinfo {author} {\bibfnamefont {V.}~\bibnamefont
  {{Paschalidis}}}, \bibinfo {author} {\bibfnamefont {W.~E.}\ \bibnamefont
  {{East}}}, \bibinfo {author} {\bibfnamefont {F.}~\bibnamefont {{Pretorius}}},
  \ and\ \bibinfo {author} {\bibfnamefont {S.~L.}\ \bibnamefont {{Shapiro}}},\
  }\href {\doibase 10.1103/PhysRevD.92.121502} {\bibfield  {journal} {\bibinfo
  {journal} {\prd}\ }\textbf {\bibinfo {volume} {92}},\ \bibinfo {eid} {121502}
  (\bibinfo {year} {2015}{\natexlab{a}})},\ \Eprint
  {http://arxiv.org/abs/1510.03432} {arXiv:1510.03432 [astro-ph.HE]}
  \BibitemShut {NoStop}%
\bibitem [{\citenamefont {{East}}\ \emph {et~al.}(2016)\citenamefont {{East}},
  \citenamefont {{Paschalidis}}, \citenamefont {{Pretorius}},\ and\
  \citenamefont {{Shapiro}}}]{EPPS2016}%
  \BibitemOpen
  \bibfield  {author} {\bibinfo {author} {\bibfnamefont {W.~E.}\ \bibnamefont
  {{East}}}, \bibinfo {author} {\bibfnamefont {V.}~\bibnamefont
  {{Paschalidis}}}, \bibinfo {author} {\bibfnamefont {F.}~\bibnamefont
  {{Pretorius}}}, \ and\ \bibinfo {author} {\bibfnamefont {S.~L.}\ \bibnamefont
  {{Shapiro}}},\ }\href {\doibase 10.1103/PhysRevD.93.024011} {\bibfield
  {journal} {\bibinfo  {journal} {\prd}\ }\textbf {\bibinfo {volume} {93}},\
  \bibinfo {eid} {024011} (\bibinfo {year} {2016})},\ \Eprint
  {http://arxiv.org/abs/1511.01093} {arXiv:1511.01093 [astro-ph.HE]}
  \BibitemShut {NoStop}%
\bibitem [{\citenamefont {East}\ \emph {et~al.}(2016)\citenamefont {East},
  \citenamefont {Paschalidis},\ and\ \citenamefont {Pretorius}}]{East:2016zvv}%
  \BibitemOpen
  \bibfield  {author} {\bibinfo {author} {\bibfnamefont {W.~E.}\ \bibnamefont
  {East}}, \bibinfo {author} {\bibfnamefont {V.}~\bibnamefont {Paschalidis}}, \
  and\ \bibinfo {author} {\bibfnamefont {F.}~\bibnamefont {Pretorius}},\ }\href
  {\doibase 10.1088/0264-9381/33/24/244004} {\bibfield  {journal} {\bibinfo
  {journal} {Class. Quant. Grav.}\ }\textbf {\bibinfo {volume} {33}},\ \bibinfo
  {pages} {244004} (\bibinfo {year} {2016})},\ \Eprint
  {http://arxiv.org/abs/1609.00725} {arXiv:1609.00725 [astro-ph.HE]}
  \BibitemShut {NoStop}%
\bibitem [{\citenamefont {Dietrich}\ \emph
  {et~al.}(2018{\natexlab{c}})\citenamefont {Dietrich}, \citenamefont
  {Bernuzzi}, \citenamefont {Brügmann}, \citenamefont {Ujevic},\ and\
  \citenamefont {Tichy}}]{Dietrich:2017xqb}%
  \BibitemOpen
  \bibfield  {author} {\bibinfo {author} {\bibfnamefont {T.}~\bibnamefont
  {Dietrich}}, \bibinfo {author} {\bibfnamefont {S.}~\bibnamefont {Bernuzzi}},
  \bibinfo {author} {\bibfnamefont {B.}~\bibnamefont {Brügmann}}, \bibinfo
  {author} {\bibfnamefont {M.}~\bibnamefont {Ujevic}}, \ and\ \bibinfo {author}
  {\bibfnamefont {W.}~\bibnamefont {Tichy}},\ }\href {\doibase
  10.1103/PhysRevD.97.064002} {\bibfield  {journal} {\bibinfo  {journal} {Phys.
  Rev.}\ }\textbf {\bibinfo {volume} {D97}},\ \bibinfo {pages} {064002}
  (\bibinfo {year} {2018}{\natexlab{c}})},\ \Eprint
  {http://arxiv.org/abs/1712.02992} {arXiv:1712.02992 [gr-qc]} \BibitemShut
  {NoStop}%
\bibitem [{\citenamefont {Ruiz}\ \emph {et~al.}(2019)\citenamefont {Ruiz},
  \citenamefont {Tsokaros}, \citenamefont {Paschalidis},\ and\ \citenamefont
  {Shapiro}}]{Ruiz:2019ezy}%
  \BibitemOpen
  \bibfield  {author} {\bibinfo {author} {\bibfnamefont {M.}~\bibnamefont
  {Ruiz}}, \bibinfo {author} {\bibfnamefont {A.}~\bibnamefont {Tsokaros}},
  \bibinfo {author} {\bibfnamefont {V.}~\bibnamefont {Paschalidis}}, \ and\
  \bibinfo {author} {\bibfnamefont {S.~L.}\ \bibnamefont {Shapiro}},\ }\href
  {\doibase 10.1103/PhysRevD.99.084032} {\bibfield  {journal} {\bibinfo
  {journal} {Phys. Rev.}\ }\textbf {\bibinfo {volume} {D99}},\ \bibinfo {pages}
  {084032} (\bibinfo {year} {2019})},\ \Eprint
  {http://arxiv.org/abs/1902.08636} {arXiv:1902.08636 [astro-ph.HE]}
  \BibitemShut {NoStop}%
\bibitem [{\citenamefont {Most}\ \emph {et~al.}(2019)\citenamefont {Most},
  \citenamefont {Papenfort}, \citenamefont {Tsokaros},\ and\ \citenamefont
  {Rezzolla}}]{Most:2019pac}%
  \BibitemOpen
  \bibfield  {author} {\bibinfo {author} {\bibfnamefont {E.~R.}\ \bibnamefont
  {Most}}, \bibinfo {author} {\bibfnamefont {L.~J.}\ \bibnamefont {Papenfort}},
  \bibinfo {author} {\bibfnamefont {A.}~\bibnamefont {Tsokaros}}, \ and\
  \bibinfo {author} {\bibfnamefont {L.}~\bibnamefont {Rezzolla}},\ }\href@noop
  {} {\  (\bibinfo {year} {2019})},\ \Eprint {http://arxiv.org/abs/1904.04220}
  {arXiv:1904.04220 [astro-ph.HE]} \BibitemShut {NoStop}%
\bibitem [{\citenamefont {Tsokaros}\ \emph {et~al.}(2015)\citenamefont
  {Tsokaros}, \citenamefont {Ury{\=u}},\ and\ \citenamefont
  {Rezzolla}}]{Tsokaros:2015fea}%
  \BibitemOpen
  \bibfield  {author} {\bibinfo {author} {\bibfnamefont {A.}~\bibnamefont
  {Tsokaros}}, \bibinfo {author} {\bibfnamefont {K.}~\bibnamefont {Ury{\=u}}},
  \ and\ \bibinfo {author} {\bibfnamefont {L.}~\bibnamefont {Rezzolla}},\
  }\href {\doibase 10.1103/PhysRevD.91.104030} {\bibfield  {journal} {\bibinfo
  {journal} {Phys. Rev.}\ }\textbf {\bibinfo {volume} {D91}},\ \bibinfo {pages}
  {104030} (\bibinfo {year} {2015})},\ \Eprint
  {http://arxiv.org/abs/1502.05674} {arXiv:1502.05674 [gr-qc]} \BibitemShut
  {NoStop}%
\bibitem [{\citenamefont {Tsokaros}\ \emph {et~al.}(2018)\citenamefont
  {Tsokaros}, \citenamefont {Uryu}, \citenamefont {Ruiz},\ and\ \citenamefont
  {Shapiro}}]{Tsokaros:2018dqs}%
  \BibitemOpen
  \bibfield  {author} {\bibinfo {author} {\bibfnamefont {A.}~\bibnamefont
  {Tsokaros}}, \bibinfo {author} {\bibfnamefont {K.}~\bibnamefont {Uryu}},
  \bibinfo {author} {\bibfnamefont {M.}~\bibnamefont {Ruiz}}, \ and\ \bibinfo
  {author} {\bibfnamefont {S.~L.}\ \bibnamefont {Shapiro}},\ }\href {\doibase
  10.1103/PhysRevD.98.124019} {\bibfield  {journal} {\bibinfo  {journal} {Phys.
  Rev.}\ }\textbf {\bibinfo {volume} {D98}},\ \bibinfo {pages} {124019}
  (\bibinfo {year} {2018})},\ \Eprint {http://arxiv.org/abs/1809.08237}
  {arXiv:1809.08237 [gr-qc]} \BibitemShut {NoStop}%
\bibitem [{\citenamefont {Tsokaros}\ \emph {et~al.}(2016)\citenamefont
  {Tsokaros}, \citenamefont {Mundim}, \citenamefont {Galeazzi}, \citenamefont
  {Rezzolla},\ and\ \citenamefont {Ury{\=u}}}]{Tsokaros:2016eik}%
  \BibitemOpen
  \bibfield  {author} {\bibinfo {author} {\bibfnamefont {A.}~\bibnamefont
  {Tsokaros}}, \bibinfo {author} {\bibfnamefont {B.~C.}\ \bibnamefont
  {Mundim}}, \bibinfo {author} {\bibfnamefont {F.}~\bibnamefont {Galeazzi}},
  \bibinfo {author} {\bibfnamefont {L.}~\bibnamefont {Rezzolla}}, \ and\
  \bibinfo {author} {\bibfnamefont {K.}~\bibnamefont {Ury{\=u}}},\ }\href
  {\doibase 10.1103/PhysRevD.94.044049} {\bibfield  {journal} {\bibinfo
  {journal} {Phys. Rev.}\ }\textbf {\bibinfo {volume} {D94}},\ \bibinfo {pages}
  {044049} (\bibinfo {year} {2016})},\ \Eprint
  {http://arxiv.org/abs/1605.07205} {arXiv:1605.07205 [gr-qc]} \BibitemShut
  {NoStop}%
\bibitem [{\citenamefont {Ury{\=u}}\ and\ \citenamefont
  {Tsokaros}(2012)}]{Uryu:2011ky}%
  \BibitemOpen
  \bibfield  {author} {\bibinfo {author} {\bibfnamefont {K.}~\bibnamefont
  {Ury{\=u}}}\ and\ \bibinfo {author} {\bibfnamefont {A.}~\bibnamefont
  {Tsokaros}},\ }\href {\doibase 10.1103/PhysRevD.85.064014} {\bibfield
  {journal} {\bibinfo  {journal} {Phys. Rev.}\ }\textbf {\bibinfo {volume}
  {D85}},\ \bibinfo {pages} {064014} (\bibinfo {year} {2012})},\ \Eprint
  {http://arxiv.org/abs/1108.3065} {arXiv:1108.3065 [gr-qc]} \BibitemShut
  {NoStop}%
\bibitem [{\citenamefont {Etienne}\ \emph {et~al.}(2012)\citenamefont
  {Etienne}, \citenamefont {Paschalidis},\ and\ \citenamefont
  {Shapiro}}]{Etienne:2012te}%
  \BibitemOpen
  \bibfield  {author} {\bibinfo {author} {\bibfnamefont {Z.~B.}\ \bibnamefont
  {Etienne}}, \bibinfo {author} {\bibfnamefont {V.}~\bibnamefont
  {Paschalidis}}, \ and\ \bibinfo {author} {\bibfnamefont {S.~L.}\ \bibnamefont
  {Shapiro}},\ }\href {\doibase 10.1103/PhysRevD.86.084026} {\bibfield
  {journal} {\bibinfo  {journal} {Phys.Rev.}\ }\textbf {\bibinfo {volume}
  {D86}},\ \bibinfo {pages} {084026} (\bibinfo {year} {2012})}\BibitemShut
  {NoStop}%
\bibitem [{\citenamefont {{Etienne}}\ \emph
  {et~al.}(2012{\natexlab{a}})\citenamefont {{Etienne}}, \citenamefont {{Liu}},
  \citenamefont {{Paschalidis}},\ and\ \citenamefont
  {{Shapiro}}}]{UIUC_PAPER1}%
  \BibitemOpen
  \bibfield  {author} {\bibinfo {author} {\bibfnamefont {Z.~B.}\ \bibnamefont
  {{Etienne}}}, \bibinfo {author} {\bibfnamefont {Y.~T.}\ \bibnamefont
  {{Liu}}}, \bibinfo {author} {\bibfnamefont {V.}~\bibnamefont
  {{Paschalidis}}}, \ and\ \bibinfo {author} {\bibfnamefont {S.~L.}\
  \bibnamefont {{Shapiro}}},\ }\href {\doibase 10.1103/PhysRevD.85.064029}
  {\bibfield  {journal} {\bibinfo  {journal} {prd}\ }\textbf {\bibinfo {volume}
  {85}},\ \bibinfo {eid} {064029} (\bibinfo {year}
  {2012}{\natexlab{a}})}\BibitemShut {NoStop}%
\bibitem [{\citenamefont {{Etienne}}\ \emph
  {et~al.}(2012{\natexlab{b}})\citenamefont {{Etienne}}, \citenamefont
  {{Paschalidis}},\ and\ \citenamefont {{Shapiro}}}]{UIUC_PAPER2}%
  \BibitemOpen
  \bibfield  {author} {\bibinfo {author} {\bibfnamefont {Z.~B.}\ \bibnamefont
  {{Etienne}}}, \bibinfo {author} {\bibfnamefont {V.}~\bibnamefont
  {{Paschalidis}}}, \ and\ \bibinfo {author} {\bibfnamefont {S.~L.}\
  \bibnamefont {{Shapiro}}},\ }\href {\doibase 10.1103/PhysRevD.86.084026}
  {\bibfield  {journal} {\bibinfo  {journal} {prd}\ }\textbf {\bibinfo {volume}
  {86}},\ \bibinfo {eid} {084026} (\bibinfo {year}
  {2012}{\natexlab{b}})}\BibitemShut {NoStop}%
\bibitem [{\citenamefont {{Paschalidis}}\ \emph
  {et~al.}(2015{\natexlab{b}})\citenamefont {{Paschalidis}}, \citenamefont
  {{Ruiz}},\ and\ \citenamefont {{Shapiro}}}]{prs15}%
  \BibitemOpen
  \bibfield  {author} {\bibinfo {author} {\bibfnamefont {V.}~\bibnamefont
  {{Paschalidis}}}, \bibinfo {author} {\bibfnamefont {M.}~\bibnamefont
  {{Ruiz}}}, \ and\ \bibinfo {author} {\bibfnamefont {S.~L.}\ \bibnamefont
  {{Shapiro}}},\ }\href {\doibase 10.1088/2041-8205/806/1/L14} {\bibfield
  {journal} {\bibinfo  {journal} {Astrophys. J. Letters}\ }\textbf {\bibinfo
  {volume} {806}},\ \bibinfo {eid} {L14} (\bibinfo {year}
  {2015}{\natexlab{b}})}\BibitemShut {NoStop}%
\bibitem [{\citenamefont {{Baumgarte}}\ \emph {et~al.}(1997)\citenamefont
  {{Baumgarte}}, \citenamefont {{Cook}}, \citenamefont {{Scheel}},
  \citenamefont {{Shapiro}},\ and\ \citenamefont
  {{Teukolsky}}}]{1997PhRvL..79.1182B}%
  \BibitemOpen
  \bibfield  {author} {\bibinfo {author} {\bibfnamefont {T.~W.}\ \bibnamefont
  {{Baumgarte}}}, \bibinfo {author} {\bibfnamefont {G.~B.}\ \bibnamefont
  {{Cook}}}, \bibinfo {author} {\bibfnamefont {M.~A.}\ \bibnamefont
  {{Scheel}}}, \bibinfo {author} {\bibfnamefont {S.~L.}\ \bibnamefont
  {{Shapiro}}}, \ and\ \bibinfo {author} {\bibfnamefont {S.~A.}\ \bibnamefont
  {{Teukolsky}}},\ }\href {\doibase 10.1103/PhysRevLett.79.1182} {\bibfield
  {journal} {\bibinfo  {journal} {\prl}\ }\textbf {\bibinfo {volume} {79}},\
  \bibinfo {pages} {1182} (\bibinfo {year} {1997})},\ \Eprint
  {http://arxiv.org/abs/gr-qc/9704024} {arXiv:gr-qc/9704024} \BibitemShut
  {NoStop}%
\bibitem [{\citenamefont {{Baumgarte}}\ \emph {et~al.}(1998)\citenamefont
  {{Baumgarte}}, \citenamefont {{Cook}}, \citenamefont {{Scheel}},
  \citenamefont {{Shapiro}},\ and\ \citenamefont
  {{Teukolsky}}}]{1998PhRvD..57.7299B}%
  \BibitemOpen
  \bibfield  {author} {\bibinfo {author} {\bibfnamefont {T.~W.}\ \bibnamefont
  {{Baumgarte}}}, \bibinfo {author} {\bibfnamefont {G.~B.}\ \bibnamefont
  {{Cook}}}, \bibinfo {author} {\bibfnamefont {M.~A.}\ \bibnamefont
  {{Scheel}}}, \bibinfo {author} {\bibfnamefont {S.~L.}\ \bibnamefont
  {{Shapiro}}}, \ and\ \bibinfo {author} {\bibfnamefont {S.~A.}\ \bibnamefont
  {{Teukolsky}}},\ }\href {\doibase 10.1103/PhysRevD.57.7299} {\bibfield
  {journal} {\bibinfo  {journal} {\prd}\ }\textbf {\bibinfo {volume} {57}},\
  \bibinfo {pages} {7299} (\bibinfo {year} {1998})},\ \Eprint
  {http://arxiv.org/abs/gr-qc/9709026} {arXiv:gr-qc/9709026} \BibitemShut
  {NoStop}%
\bibitem [{\citenamefont {{Marronetti}}\ \emph {et~al.}(1998)\citenamefont
  {{Marronetti}}, \citenamefont {{Mathews}},\ and\ \citenamefont
  {{Wilson}}}]{1998PhRvD..58j7503M}%
  \BibitemOpen
  \bibfield  {author} {\bibinfo {author} {\bibfnamefont {P.}~\bibnamefont
  {{Marronetti}}}, \bibinfo {author} {\bibfnamefont {G.~J.}\ \bibnamefont
  {{Mathews}}}, \ and\ \bibinfo {author} {\bibfnamefont {J.~R.}\ \bibnamefont
  {{Wilson}}},\ }\href {\doibase 10.1103/PhysRevD.58.107503} {\bibfield
  {journal} {\bibinfo  {journal} {\prd}\ }\textbf {\bibinfo {volume} {58}},\
  \bibinfo {eid} {107503} (\bibinfo {year} {1998})},\ \Eprint
  {http://arxiv.org/abs/gr-qc/9803093} {arXiv:gr-qc/9803093} \BibitemShut
  {NoStop}%
\bibitem [{\citenamefont {{Bildsten}}\ and\ \citenamefont
  {{Cutler}}(1992)}]{1992ApJ...400..175B}%
  \BibitemOpen
  \bibfield  {author} {\bibinfo {author} {\bibfnamefont {L.}~\bibnamefont
  {{Bildsten}}}\ and\ \bibinfo {author} {\bibfnamefont {C.}~\bibnamefont
  {{Cutler}}},\ }\href {\doibase 10.1086/171983} {\bibfield  {journal}
  {\bibinfo  {journal} {\apj}\ }\textbf {\bibinfo {volume} {400}},\ \bibinfo
  {pages} {175} (\bibinfo {year} {1992})}\BibitemShut {NoStop}%
\bibitem [{\citenamefont {{Kochanek}}(1992)}]{1992ApJ...398..234K}%
  \BibitemOpen
  \bibfield  {author} {\bibinfo {author} {\bibfnamefont {C.~S.}\ \bibnamefont
  {{Kochanek}}},\ }\href {\doibase 10.1086/171851} {\bibfield  {journal}
  {\bibinfo  {journal} {\apj}\ }\textbf {\bibinfo {volume} {398}},\ \bibinfo
  {pages} {234} (\bibinfo {year} {1992})}\BibitemShut {NoStop}%
\bibitem [{\citenamefont {{Alford}}\ \emph {et~al.}(2005)\citenamefont
  {{Alford}}, \citenamefont {{Braby}}, \citenamefont {{Paris}},\ and\
  \citenamefont {{Reddy}}}]{Alford2005}%
  \BibitemOpen
  \bibfield  {author} {\bibinfo {author} {\bibfnamefont {M.}~\bibnamefont
  {{Alford}}}, \bibinfo {author} {\bibfnamefont {M.}~\bibnamefont {{Braby}}},
  \bibinfo {author} {\bibfnamefont {M.}~\bibnamefont {{Paris}}}, \ and\
  \bibinfo {author} {\bibfnamefont {S.}~\bibnamefont {{Reddy}}},\ }\href
  {\doibase 10.1086/430902} {\bibfield  {journal} {\bibinfo  {journal}
  {Astrophys. J.}\ }\textbf {\bibinfo {volume} {629}},\ \bibinfo {pages} {969}
  (\bibinfo {year} {2005})},\ \Eprint {http://arxiv.org/abs/nucl-th/0411016}
  {nucl-th/0411016} \BibitemShut {NoStop}%
\bibitem [{\citenamefont {{Akmal}}\ \emph {et~al.}(1998)\citenamefont
  {{Akmal}}, \citenamefont {{Pandharipande}},\ and\ \citenamefont
  {{Ravenhall}}}]{Akmal1998a}%
  \BibitemOpen
  \bibfield  {author} {\bibinfo {author} {\bibfnamefont {A.}~\bibnamefont
  {{Akmal}}}, \bibinfo {author} {\bibfnamefont {V.~R.}\ \bibnamefont
  {{Pandharipande}}}, \ and\ \bibinfo {author} {\bibfnamefont {D.~G.}\
  \bibnamefont {{Ravenhall}}},\ }\href {\doibase 10.1103/PhysRevC.58.1804}
  {\bibfield  {journal} {\bibinfo  {journal} {Phys. Rev. C}\ }\textbf {\bibinfo
  {volume} {58}},\ \bibinfo {pages} {1804} (\bibinfo {year} {1998})},\ \Eprint
  {http://arxiv.org/abs/arXiv:hep-ph/9804388} {arXiv:hep-ph/9804388}
  \BibitemShut {NoStop}%
\bibitem [{\citenamefont {{Douchin}}\ and\ \citenamefont
  {{Haensel}}(2001)}]{Douchin01}%
  \BibitemOpen
  \bibfield  {author} {\bibinfo {author} {\bibfnamefont {F.}~\bibnamefont
  {{Douchin}}}\ and\ \bibinfo {author} {\bibfnamefont {P.}~\bibnamefont
  {{Haensel}}},\ }\href {\doibase 10.1051/0004-6361:20011402} {\bibfield
  {journal} {\bibinfo  {journal} {Astron. Astrophys.}\ }\textbf {\bibinfo
  {volume} {380}},\ \bibinfo {pages} {151} (\bibinfo {year} {2001})},\ \Eprint
  {http://arxiv.org/abs/arXiv:astro-ph/0111092} {arXiv:astro-ph/0111092}
  \BibitemShut {NoStop}%
\bibitem [{\citenamefont {Hinderer}(2008)}]{Hinderer:2007mb}%
  \BibitemOpen
  \bibfield  {author} {\bibinfo {author} {\bibfnamefont {T.}~\bibnamefont
  {Hinderer}},\ }\href {\doibase 10.1086/533487} {\bibfield  {journal}
  {\bibinfo  {journal} {Astrophys. J.}\ }\textbf {\bibinfo {volume} {677}},\
  \bibinfo {pages} {1216} (\bibinfo {year} {2008})},\ \Eprint
  {http://arxiv.org/abs/0711.2420} {arXiv:0711.2420 [astro-ph]} \BibitemShut
  {NoStop}%
\bibitem [{\citenamefont {{Annala}}\ \emph {et~al.}(2018)\citenamefont
  {{Annala}}, \citenamefont {{Gorda}}, \citenamefont {{Kurkela}},\ and\
  \citenamefont {{Vuorinen}}}]{Annala2017}%
  \BibitemOpen
  \bibfield  {author} {\bibinfo {author} {\bibfnamefont {E.}~\bibnamefont
  {{Annala}}}, \bibinfo {author} {\bibfnamefont {T.}~\bibnamefont {{Gorda}}},
  \bibinfo {author} {\bibfnamefont {A.}~\bibnamefont {{Kurkela}}}, \ and\
  \bibinfo {author} {\bibfnamefont {A.}~\bibnamefont {{Vuorinen}}},\ }\href
  {\doibase 10.1103/PhysRevLett.120.172703} {\bibfield  {journal} {\bibinfo
  {journal} {Phys. Rev. Lett.}\ }\textbf {\bibinfo {volume} {120}},\ \bibinfo
  {eid} {172703} (\bibinfo {year} {2018})},\ \Eprint
  {http://arxiv.org/abs/1711.02644} {arXiv:1711.02644 [astro-ph.HE]}
  \BibitemShut {NoStop}%
\bibitem [{\citenamefont {{Bauswein}}\ \emph {et~al.}(2017)\citenamefont
  {{Bauswein}}, \citenamefont {{Just}}, \citenamefont {{Janka}},\ and\
  \citenamefont {{Stergioulas}}}]{Bauswein2017b}%
  \BibitemOpen
  \bibfield  {author} {\bibinfo {author} {\bibfnamefont {A.}~\bibnamefont
  {{Bauswein}}}, \bibinfo {author} {\bibfnamefont {O.}~\bibnamefont {{Just}}},
  \bibinfo {author} {\bibfnamefont {H.-T.}\ \bibnamefont {{Janka}}}, \ and\
  \bibinfo {author} {\bibfnamefont {N.}~\bibnamefont {{Stergioulas}}},\ }\href
  {\doibase 10.3847/2041-8213/aa9994} {\bibfield  {journal} {\bibinfo
  {journal} {Astrophys. J. Lett.}\ }\textbf {\bibinfo {volume} {850}},\
  \bibinfo {eid} {L34} (\bibinfo {year} {2017})},\ \Eprint
  {http://arxiv.org/abs/1710.06843} {arXiv:1710.06843 [astro-ph.HE]}
  \BibitemShut {NoStop}%
\bibitem [{\citenamefont {{Radice}}\ \emph {et~al.}(2018)\citenamefont
  {{Radice}}, \citenamefont {{Perego}}, \citenamefont {{Zappa}},\ and\
  \citenamefont {{Bernuzzi}}}]{Radice2017b}%
  \BibitemOpen
  \bibfield  {author} {\bibinfo {author} {\bibfnamefont {D.}~\bibnamefont
  {{Radice}}}, \bibinfo {author} {\bibfnamefont {A.}~\bibnamefont {{Perego}}},
  \bibinfo {author} {\bibfnamefont {F.}~\bibnamefont {{Zappa}}}, \ and\
  \bibinfo {author} {\bibfnamefont {S.}~\bibnamefont {{Bernuzzi}}},\ }\href
  {\doibase 10.3847/2041-8213/aaa402} {\bibfield  {journal} {\bibinfo
  {journal} {Astrophys. J. Lett.}\ }\textbf {\bibinfo {volume} {852}},\
  \bibinfo {eid} {L29} (\bibinfo {year} {2018})},\ \Eprint
  {http://arxiv.org/abs/1711.03647} {arXiv:1711.03647 [astro-ph.HE]}
  \BibitemShut {NoStop}%
\bibitem [{\citenamefont {{Abbott}}\ \emph {et~al.}(2018)\citenamefont
  {{Abbott}}, \citenamefont {{Abbott}}, \citenamefont {{Abbott}}, \citenamefont
  {{Acernese}}, \citenamefont {{Ackley}}, \citenamefont {{Adams}},
  \citenamefont {{Adams}}, \citenamefont {{Addesso}}, \citenamefont
  {{Adhikari}}, \citenamefont {{Adya}},\ and\ \citenamefont
  {et~al.}}]{Abbott2018b}%
  \BibitemOpen
  \bibfield  {author} {\bibinfo {author} {\bibfnamefont {B.~P.}\ \bibnamefont
  {{Abbott}}}, \bibinfo {author} {\bibfnamefont {R.}~\bibnamefont {{Abbott}}},
  \bibinfo {author} {\bibfnamefont {T.~D.}\ \bibnamefont {{Abbott}}}, \bibinfo
  {author} {\bibfnamefont {F.}~\bibnamefont {{Acernese}}}, \bibinfo {author}
  {\bibfnamefont {K.}~\bibnamefont {{Ackley}}}, \bibinfo {author}
  {\bibfnamefont {C.}~\bibnamefont {{Adams}}}, \bibinfo {author} {\bibfnamefont
  {T.}~\bibnamefont {{Adams}}}, \bibinfo {author} {\bibfnamefont
  {P.}~\bibnamefont {{Addesso}}}, \bibinfo {author} {\bibfnamefont {R.~X.}\
  \bibnamefont {{Adhikari}}}, \bibinfo {author} {\bibfnamefont {V.~B.}\
  \bibnamefont {{Adya}}}, \ and\ \bibinfo {author} {\bibnamefont {et~al.}},\
  }\href {\doibase 10.1103/PhysRevLett.121.161101} {\bibfield  {journal}
  {\bibinfo  {journal} {Physical Review Letters}\ }\textbf {\bibinfo {volume}
  {121}},\ \bibinfo {eid} {161101} (\bibinfo {year} {2018})},\ \Eprint
  {http://arxiv.org/abs/1805.11581} {arXiv:1805.11581 [gr-qc]} \BibitemShut
  {NoStop}%
\bibitem [{\citenamefont {{Most}}\ \emph {et~al.}(2018)\citenamefont {{Most}},
  \citenamefont {{Weih}}, \citenamefont {{Rezzolla}},\ and\ \citenamefont
  {{Schaffner-Bielich}}}]{Most2018}%
  \BibitemOpen
  \bibfield  {author} {\bibinfo {author} {\bibfnamefont {E.~R.}\ \bibnamefont
  {{Most}}}, \bibinfo {author} {\bibfnamefont {L.~R.}\ \bibnamefont {{Weih}}},
  \bibinfo {author} {\bibfnamefont {L.}~\bibnamefont {{Rezzolla}}}, \ and\
  \bibinfo {author} {\bibfnamefont {J.}~\bibnamefont {{Schaffner-Bielich}}},\
  }\href {\doibase 10.1103/PhysRevLett.120.261103} {\bibfield  {journal}
  {\bibinfo  {journal} {Phys. Rev. Lett.}\ }\textbf {\bibinfo {volume} {120}},\
  \bibinfo {eid} {261103} (\bibinfo {year} {2018})},\ \Eprint
  {http://arxiv.org/abs/1803.00549} {arXiv:1803.00549 [gr-qc]} \BibitemShut
  {NoStop}%
\bibitem [{\citenamefont {Kiuchi}\ \emph {et~al.}(2019)\citenamefont {Kiuchi},
  \citenamefont {Kyutoku}, \citenamefont {Shibata},\ and\ \citenamefont
  {Taniguchi}}]{Kiuchi2019}%
  \BibitemOpen
  \bibfield  {author} {\bibinfo {author} {\bibfnamefont {K.}~\bibnamefont
  {Kiuchi}}, \bibinfo {author} {\bibfnamefont {K.}~\bibnamefont {Kyutoku}},
  \bibinfo {author} {\bibfnamefont {M.}~\bibnamefont {Shibata}}, \ and\
  \bibinfo {author} {\bibfnamefont {K.}~\bibnamefont {Taniguchi}},\ }\href@noop
  {} {\  (\bibinfo {year} {2019})},\ \Eprint {http://arxiv.org/abs/1903.01466}
  {arXiv:1903.01466 [astro-ph.HE]} \BibitemShut {NoStop}%
\bibitem [{\citenamefont {Allen}\ \emph {et~al.}(2001)\citenamefont {Allen},
  \citenamefont {Angulo}, \citenamefont {Foster}, \citenamefont {Lanfermann},
  \citenamefont {Liu}, \citenamefont {Radke}, \citenamefont {Seidel},\ and\
  \citenamefont {Shalf}}]{AllAngFos01}%
  \BibitemOpen
  \bibfield  {author} {\bibinfo {author} {\bibfnamefont {G.}~\bibnamefont
  {Allen}}, \bibinfo {author} {\bibfnamefont {D.}~\bibnamefont {Angulo}},
  \bibinfo {author} {\bibfnamefont {I.}~\bibnamefont {Foster}}, \bibinfo
  {author} {\bibfnamefont {G.}~\bibnamefont {Lanfermann}}, \bibinfo {author}
  {\bibfnamefont {C.}~\bibnamefont {Liu}}, \bibinfo {author} {\bibfnamefont
  {T.}~\bibnamefont {Radke}}, \bibinfo {author} {\bibfnamefont
  {E.}~\bibnamefont {Seidel}}, \ and\ \bibinfo {author} {\bibfnamefont
  {J.}~\bibnamefont {Shalf}},\ }\href@noop {} {\bibfield  {journal} {\bibinfo
  {journal} {Int. J. of High Performance Computing Applications}\ }\textbf
  {\bibinfo {volume} {15}} (\bibinfo {year} {2001})}\BibitemShut {NoStop}%
\bibitem [{Cactus()}]{cactusweb}%
  \BibitemOpen
  Cactus,\ \href {http://cactuscode.org/} {\enquote {\bibinfo {title}
  {Cactuscode, http://cactuscode.org/},}\ }\BibitemShut {NoStop}%
\bibitem [{\citenamefont {Schnetter}\ \emph {et~al.}(2004)\citenamefont
  {Schnetter}, \citenamefont {Hawley},\ and\ \citenamefont {Hawke}}]{Carpet}%
  \BibitemOpen
  \bibfield  {author} {\bibinfo {author} {\bibfnamefont {E.}~\bibnamefont
  {Schnetter}}, \bibinfo {author} {\bibfnamefont {S.~H.}\ \bibnamefont
  {Hawley}}, \ and\ \bibinfo {author} {\bibfnamefont {I.}~\bibnamefont
  {Hawke}},\ }\href {\doibase 10.1088/0264-9381/21/6/014} {\bibfield  {journal}
  {\bibinfo  {journal} {Class. Quantum Grav.}\ }\textbf {\bibinfo {volume}
  {21}},\ \bibinfo {pages} {1465} (\bibinfo {year} {2004})},\ \Eprint
  {http://arxiv.org/abs/arXiv:gr-qc/0310042} {arXiv:gr-qc/0310042} \BibitemShut
  {NoStop}%
\bibitem [{Carpet()}]{carpetweb}%
  \BibitemOpen
  Carpet,\ \href {{http://www.carpetcode.org/}} {}\bibinfo {note} {{Carpet Code
  homepage}}\BibitemShut {NoStop}%
\bibitem [{\citenamefont {Shibata}\ and\ \citenamefont
  {Nakamura}(1995)}]{shibnak95}%
  \BibitemOpen
  \bibfield  {author} {\bibinfo {author} {\bibfnamefont {M.}~\bibnamefont
  {Shibata}}\ and\ \bibinfo {author} {\bibfnamefont {T.}~\bibnamefont
  {Nakamura}},\ }\href {\doibase 10.1103/PhysRevD.52.5428} {\bibfield
  {journal} {\bibinfo  {journal} {Phys. Rev. D}\ }\textbf {\bibinfo {volume}
  {52}},\ \bibinfo {pages} {5428} (\bibinfo {year} {1995})}\BibitemShut
  {NoStop}%
\bibitem [{\citenamefont {{Baumgarte}}\ and\ \citenamefont
  {{Shapiro}}(1998)}]{BS}%
  \BibitemOpen
  \bibfield  {author} {\bibinfo {author} {\bibfnamefont {T.~W.}\ \bibnamefont
  {{Baumgarte}}}\ and\ \bibinfo {author} {\bibfnamefont {S.~L.}\ \bibnamefont
  {{Shapiro}}},\ }\href@noop {} {\bibfield  {journal} {\bibinfo  {journal}
  {prd}\ }\textbf {\bibinfo {volume} {59}},\ \bibinfo {pages} {024007}
  (\bibinfo {year} {1998})}\BibitemShut {NoStop}%
\bibitem [{\citenamefont {{Baumgarte}}\ and\ \citenamefont
  {{Shapiro}}(2010)}]{BSBook}%
  \BibitemOpen
  \bibfield  {author} {\bibinfo {author} {\bibfnamefont {T.~W.}\ \bibnamefont
  {{Baumgarte}}}\ and\ \bibinfo {author} {\bibfnamefont {S.~L.}\ \bibnamefont
  {{Shapiro}}},\ }\href@noop {} {\emph {\bibinfo {title} {{Numerical
  Relativity: Solving Einstein's Equations on the Computer}}}}\ (\bibinfo
  {publisher} {Cambridge University Press},\ \bibinfo {year}
  {2010})\BibitemShut {NoStop}%
\bibitem [{\citenamefont {Etienne}\ \emph {et~al.}(2008)\citenamefont
  {Etienne}, \citenamefont {Faber}, \citenamefont {Liu}, \citenamefont
  {Shapiro}, \citenamefont {Taniguchi} \emph {et~al.}}]{Etienne:2007jg}%
  \BibitemOpen
  \bibfield  {author} {\bibinfo {author} {\bibfnamefont {Z.~B.}\ \bibnamefont
  {Etienne}}, \bibinfo {author} {\bibfnamefont {J.~A.}\ \bibnamefont {Faber}},
  \bibinfo {author} {\bibfnamefont {Y.~T.}\ \bibnamefont {Liu}}, \bibinfo
  {author} {\bibfnamefont {S.~L.}\ \bibnamefont {Shapiro}}, \bibinfo {author}
  {\bibfnamefont {K.}~\bibnamefont {Taniguchi}},  \emph {et~al.},\ }\href
  {\doibase 10.1103/PhysRevD.77.084002} {\bibfield  {journal} {\bibinfo
  {journal} {Phys.Rev.}\ }\textbf {\bibinfo {volume} {D77}},\ \bibinfo {pages}
  {084002} (\bibinfo {year} {2008})}\BibitemShut {NoStop}%
\bibitem [{\citenamefont {{Duez}}\ \emph {et~al.}(2003)\citenamefont {{Duez}},
  \citenamefont {{Marronetti}}, \citenamefont {{Shapiro}},\ and\ \citenamefont
  {{Baumgarte}}}]{DMSB}%
  \BibitemOpen
  \bibfield  {author} {\bibinfo {author} {\bibfnamefont {M.~D.}\ \bibnamefont
  {{Duez}}}, \bibinfo {author} {\bibfnamefont {P.}~\bibnamefont
  {{Marronetti}}}, \bibinfo {author} {\bibfnamefont {S.~L.}\ \bibnamefont
  {{Shapiro}}}, \ and\ \bibinfo {author} {\bibfnamefont {T.~W.}\ \bibnamefont
  {{Baumgarte}}},\ }\href@noop {} {\bibfield  {journal} {\bibinfo  {journal}
  {prd}\ }\textbf {\bibinfo {volume} {67}},\ \bibinfo {pages} {024004}
  (\bibinfo {year} {2003})}\BibitemShut {NoStop}%
\bibitem [{\citenamefont {{Baker}}\ \emph {et~al.}(2006)\citenamefont
  {{Baker}}, \citenamefont {{Centrella}}, \citenamefont {{Choi}}, \citenamefont
  {{Koppitz}},\ and\ \citenamefont {{van Meter}}}]{goddard06}%
  \BibitemOpen
  \bibfield  {author} {\bibinfo {author} {\bibfnamefont {J.~G.}\ \bibnamefont
  {{Baker}}}, \bibinfo {author} {\bibfnamefont {J.}~\bibnamefont
  {{Centrella}}}, \bibinfo {author} {\bibfnamefont {D.-I.}\ \bibnamefont
  {{Choi}}}, \bibinfo {author} {\bibfnamefont {M.}~\bibnamefont {{Koppitz}}}, \
  and\ \bibinfo {author} {\bibfnamefont {J.}~\bibnamefont {{van Meter}}},\
  }\href {\doibase 10.1103/PhysRevD.73.104002} {\bibfield  {journal} {\bibinfo
  {journal} {prd}\ }\textbf {\bibinfo {volume} {73}},\ \bibinfo {pages}
  {104002} (\bibinfo {year} {2006})}\BibitemShut {NoStop}%
\bibitem [{\citenamefont {Etienne}\ \emph {et~al.}(2012)\citenamefont
  {Etienne}, \citenamefont {Paschalidis}, \citenamefont {Liu},\ and\
  \citenamefont {Shapiro}}]{Etienne:2011re}%
  \BibitemOpen
  \bibfield  {author} {\bibinfo {author} {\bibfnamefont {Z.~B.}\ \bibnamefont
  {Etienne}}, \bibinfo {author} {\bibfnamefont {V.}~\bibnamefont
  {Paschalidis}}, \bibinfo {author} {\bibfnamefont {Y.~T.}\ \bibnamefont
  {Liu}}, \ and\ \bibinfo {author} {\bibfnamefont {S.~L.}\ \bibnamefont
  {Shapiro}},\ }\href {\doibase 10.1103/PhysRevD.85.024013} {\bibfield
  {journal} {\bibinfo  {journal} {Phys.Rev.}\ }\textbf {\bibinfo {volume}
  {D85}},\ \bibinfo {pages} {024013} (\bibinfo {year} {2012})}\BibitemShut
  {NoStop}%
\bibitem [{\citenamefont {Etienne}\ \emph {et~al.}(2010)\citenamefont
  {Etienne}, \citenamefont {Liu},\ and\ \citenamefont
  {Shapiro}}]{Etienne:2010ui}%
  \BibitemOpen
  \bibfield  {author} {\bibinfo {author} {\bibfnamefont {Z.~B.}\ \bibnamefont
  {Etienne}}, \bibinfo {author} {\bibfnamefont {Y.~T.}\ \bibnamefont {Liu}}, \
  and\ \bibinfo {author} {\bibfnamefont {S.~L.}\ \bibnamefont {Shapiro}},\
  }\href {\doibase 10.1103/PhysRevD.82.084031} {\bibfield  {journal} {\bibinfo
  {journal} {Phys.Rev.}\ }\textbf {\bibinfo {volume} {D82}},\ \bibinfo {pages}
  {084031} (\bibinfo {year} {2010})}\BibitemShut {NoStop}%
\bibitem [{\citenamefont {{Paschalidis}}\ \emph
  {et~al.}(2011{\natexlab{a}})\citenamefont {{Paschalidis}}, \citenamefont
  {{Liu}}, \citenamefont {{Etienne}},\ and\ \citenamefont
  {{Shapiro}}}]{2011PhRvD..84j4032P}%
  \BibitemOpen
  \bibfield  {author} {\bibinfo {author} {\bibfnamefont {V.}~\bibnamefont
  {{Paschalidis}}}, \bibinfo {author} {\bibfnamefont {Y.~T.}\ \bibnamefont
  {{Liu}}}, \bibinfo {author} {\bibfnamefont {Z.}~\bibnamefont {{Etienne}}}, \
  and\ \bibinfo {author} {\bibfnamefont {S.~L.}\ \bibnamefont {{Shapiro}}},\
  }\href {\doibase 10.1103/PhysRevD.84.104032} {\bibfield  {journal} {\bibinfo
  {journal} {\prd}\ }\textbf {\bibinfo {volume} {84}},\ \bibinfo {eid} {104032}
  (\bibinfo {year} {2011}{\natexlab{a}})},\ \Eprint
  {http://arxiv.org/abs/1109.5177} {arXiv:1109.5177 [astro-ph.HE]} \BibitemShut
  {NoStop}%
\bibitem [{\citenamefont {{Paschalidis}}\ \emph
  {et~al.}(2011{\natexlab{b}})\citenamefont {{Paschalidis}}, \citenamefont
  {{Etienne}}, \citenamefont {{Liu}},\ and\ \citenamefont
  {{Shapiro}}}]{2011PhRvD..83f4002P}%
  \BibitemOpen
  \bibfield  {author} {\bibinfo {author} {\bibfnamefont {V.}~\bibnamefont
  {{Paschalidis}}}, \bibinfo {author} {\bibfnamefont {Z.}~\bibnamefont
  {{Etienne}}}, \bibinfo {author} {\bibfnamefont {Y.~T.}\ \bibnamefont
  {{Liu}}}, \ and\ \bibinfo {author} {\bibfnamefont {S.~L.}\ \bibnamefont
  {{Shapiro}}},\ }\href {\doibase 10.1103/PhysRevD.83.064002} {\bibfield
  {journal} {\bibinfo  {journal} {\prd}\ }\textbf {\bibinfo {volume} {83}},\
  \bibinfo {eid} {064002} (\bibinfo {year} {2011}{\natexlab{b}})},\ \Eprint
  {http://arxiv.org/abs/1009.4932} {arXiv:1009.4932 [astro-ph.HE]} \BibitemShut
  {NoStop}%
\bibitem [{\citenamefont {Dietrich}\ \emph
  {et~al.}(2017{\natexlab{c}})\citenamefont {Dietrich}, \citenamefont
  {Bernuzzi}, \citenamefont {Ujevic},\ and\ \citenamefont
  {Tichy}}]{Dietrich:2016lyp}%
  \BibitemOpen
  \bibfield  {author} {\bibinfo {author} {\bibfnamefont {T.}~\bibnamefont
  {Dietrich}}, \bibinfo {author} {\bibfnamefont {S.}~\bibnamefont {Bernuzzi}},
  \bibinfo {author} {\bibfnamefont {M.}~\bibnamefont {Ujevic}}, \ and\ \bibinfo
  {author} {\bibfnamefont {W.}~\bibnamefont {Tichy}},\ }\href {\doibase
  10.1103/PhysRevD.95.044045} {\bibfield  {journal} {\bibinfo  {journal} {Phys.
  Rev.}\ }\textbf {\bibinfo {volume} {D95}},\ \bibinfo {pages} {044045}
  (\bibinfo {year} {2017}{\natexlab{c}})},\ \Eprint
  {http://arxiv.org/abs/1611.07367} {arXiv:1611.07367 [gr-qc]} \BibitemShut
  {NoStop}%
\bibitem [{\citenamefont {Radice}\ \emph {et~al.}(2014)\citenamefont {Radice},
  \citenamefont {Rezzolla},\ and\ \citenamefont {Galeazzi}}]{Radice:2013hxh}%
  \BibitemOpen
  \bibfield  {author} {\bibinfo {author} {\bibfnamefont {D.}~\bibnamefont
  {Radice}}, \bibinfo {author} {\bibfnamefont {L.}~\bibnamefont {Rezzolla}}, \
  and\ \bibinfo {author} {\bibfnamefont {F.}~\bibnamefont {Galeazzi}},\ }\href
  {\doibase 10.1093/mnrasl/slt137} {\bibfield  {journal} {\bibinfo  {journal}
  {Mon. Not. Roy. Astron. Soc.}\ }\textbf {\bibinfo {volume} {437}},\ \bibinfo
  {pages} {L46} (\bibinfo {year} {2014})},\ \Eprint
  {http://arxiv.org/abs/1306.6052} {arXiv:1306.6052 [gr-qc]} \BibitemShut
  {NoStop}%
\bibitem [{\citenamefont {{Radice}}\ \emph {et~al.}(2014)\citenamefont
  {{Radice}}, \citenamefont {{Rezzolla}},\ and\ \citenamefont
  {{Galeazzi}}}]{Radice2013c}%
  \BibitemOpen
  \bibfield  {author} {\bibinfo {author} {\bibfnamefont {D.}~\bibnamefont
  {{Radice}}}, \bibinfo {author} {\bibfnamefont {L.}~\bibnamefont
  {{Rezzolla}}}, \ and\ \bibinfo {author} {\bibfnamefont {F.}~\bibnamefont
  {{Galeazzi}}},\ }\href {\doibase 10.1088/0264-9381/31/7/075012} {\bibfield
  {journal} {\bibinfo  {journal} {Class. Quantum Grav.}\ }\textbf {\bibinfo
  {volume} {31}},\ \bibinfo {eid} {075012} (\bibinfo {year} {2014})},\ \Eprint
  {http://arxiv.org/abs/1312.5004} {arXiv:1312.5004 [gr-qc]} \BibitemShut
  {NoStop}%
\bibitem [{\citenamefont {Maggiore}(2007)}]{Maggiore2007}%
  \BibitemOpen
  \bibfield  {author} {\bibinfo {author} {\bibfnamefont {M.}~\bibnamefont
  {Maggiore}},\ }\href {http://books.google.de/books?id=AqVpQgAACAAJ} {\emph
  {\bibinfo {title} {Gravitational Waves: Volume 1: Theory and Experiments}}},\
  Gravitational Waves\ (\bibinfo  {publisher} {Oxford University Press, USA},\
  \bibinfo {year} {2007})\BibitemShut {NoStop}%
\bibitem [{\citenamefont {Ruiz}\ \emph {et~al.}(2008)\citenamefont {Ruiz},
  \citenamefont {Takahashi}, \citenamefont {Alcubierre},\ and\ \citenamefont
  {Nunez}}]{Ruiz:2007yx}%
  \BibitemOpen
  \bibfield  {author} {\bibinfo {author} {\bibfnamefont {M.}~\bibnamefont
  {Ruiz}}, \bibinfo {author} {\bibfnamefont {R.}~\bibnamefont {Takahashi}},
  \bibinfo {author} {\bibfnamefont {M.}~\bibnamefont {Alcubierre}}, \ and\
  \bibinfo {author} {\bibfnamefont {D.}~\bibnamefont {Nunez}},\ }\href
  {\doibase 10.1007/s10714-007-0570-8, 10.1007/s10714-008-0684-7} {\bibfield
  {journal} {\bibinfo  {journal} {Gen. Rel. Grav.}\ }\textbf {\bibinfo {volume}
  {40}},\ \bibinfo {pages} {2467} (\bibinfo {year} {2008})},\ \Eprint
  {http://arxiv.org/abs/0707.4654} {arXiv:0707.4654 [gr-qc]} \BibitemShut
  {NoStop}%
\bibitem [{\citenamefont {Reisswig}\ and\ \citenamefont
  {Pollney}(2011)}]{Reisswig:2010di}%
  \BibitemOpen
  \bibfield  {author} {\bibinfo {author} {\bibfnamefont {C.}~\bibnamefont
  {Reisswig}}\ and\ \bibinfo {author} {\bibfnamefont {D.}~\bibnamefont
  {Pollney}},\ }\href {\doibase 10.1088/0264-9381/28/19/195015} {\bibfield
  {journal} {\bibinfo  {journal} {Class. Quant. Grav.}\ }\textbf {\bibinfo
  {volume} {28}},\ \bibinfo {pages} {195015} (\bibinfo {year} {2011})},\
  \Eprint {http://arxiv.org/abs/1006.1632} {arXiv:1006.1632 [gr-qc]}
  \BibitemShut {NoStop}%
\bibitem [{\citenamefont {{Campanelli}}\ \emph {et~al.}(2006)\citenamefont
  {{Campanelli}}, \citenamefont {{Lousto}},\ and\ \citenamefont
  {{Zlochower}}}]{clz06}%
  \BibitemOpen
  \bibfield  {author} {\bibinfo {author} {\bibfnamefont {M.}~\bibnamefont
  {{Campanelli}}}, \bibinfo {author} {\bibfnamefont {C.~O.}\ \bibnamefont
  {{Lousto}}}, \ and\ \bibinfo {author} {\bibfnamefont {Y.}~\bibnamefont
  {{Zlochower}}},\ }\href {\doibase 10.1103/PhysRevD.74.041501} {\bibfield
  {journal} {\bibinfo  {journal} {prd}\ }\textbf {\bibinfo {volume} {74}},\
  \bibinfo {pages} {041501} (\bibinfo {year} {2006})}\BibitemShut {NoStop}%
\bibitem [{\citenamefont {Tsatsin}\ and\ \citenamefont
  {Marronetti}(2013)}]{Tsatsin2013}%
  \BibitemOpen
  \bibfield  {author} {\bibinfo {author} {\bibfnamefont {P.}~\bibnamefont
  {Tsatsin}}\ and\ \bibinfo {author} {\bibfnamefont {P.}~\bibnamefont
  {Marronetti}},\ }\href {\doibase 10.1103/PhysRevD.88.064060} {\bibfield
  {journal} {\bibinfo  {journal} {Phys. Rev. D}\ }\textbf {\bibinfo {volume}
  {88}},\ \bibinfo {pages} {064060} (\bibinfo {year} {2013})}\BibitemShut
  {NoStop}%
\bibitem [{\citenamefont {Read}\ \emph
  {et~al.}(2013{\natexlab{b}})\citenamefont {Read}, \citenamefont {Baiotti},
  \citenamefont {Creighton}, \citenamefont {Friedman}, \citenamefont
  {Giacomazzo} \emph {et~al.}}]{Read:2013zra}%
  \BibitemOpen
  \bibfield  {author} {\bibinfo {author} {\bibfnamefont {J.~S.}\ \bibnamefont
  {Read}}, \bibinfo {author} {\bibfnamefont {L.}~\bibnamefont {Baiotti}},
  \bibinfo {author} {\bibfnamefont {J.~D.~E.}\ \bibnamefont {Creighton}},
  \bibinfo {author} {\bibfnamefont {J.~L.}\ \bibnamefont {Friedman}}, \bibinfo
  {author} {\bibfnamefont {B.}~\bibnamefont {Giacomazzo}},  \emph {et~al.},\
  }\href {\doibase 10.1103/PhysRevD.88.044042} {\bibfield  {journal} {\bibinfo
  {journal} {Phys.Rev.}\ }\textbf {\bibinfo {volume} {D88}},\ \bibinfo {pages}
  {044042} (\bibinfo {year} {2013}{\natexlab{b}})}\BibitemShut {NoStop}%
\bibitem [{\citenamefont {Bernuzzi}\ \emph
  {et~al.}(2014{\natexlab{b}})\citenamefont {Bernuzzi}, \citenamefont {Nagar},
  \citenamefont {Balmelli}, \citenamefont {Dietrich},\ and\ \citenamefont
  {Ujevic}}]{Bernuzzi:2014kca}%
  \BibitemOpen
  \bibfield  {author} {\bibinfo {author} {\bibfnamefont {S.}~\bibnamefont
  {Bernuzzi}}, \bibinfo {author} {\bibfnamefont {A.}~\bibnamefont {Nagar}},
  \bibinfo {author} {\bibfnamefont {S.}~\bibnamefont {Balmelli}}, \bibinfo
  {author} {\bibfnamefont {T.}~\bibnamefont {Dietrich}}, \ and\ \bibinfo
  {author} {\bibfnamefont {M.}~\bibnamefont {Ujevic}},\ }\href {\doibase
  10.1103/PhysRevLett.112.201101} {\bibfield  {journal} {\bibinfo  {journal}
  {Phys. Rev. Lett.}\ }\textbf {\bibinfo {volume} {112}},\ \bibinfo {pages}
  {201101} (\bibinfo {year} {2014}{\natexlab{b}})},\ \Eprint
  {http://arxiv.org/abs/1402.6244} {arXiv:1402.6244 [gr-qc]} \BibitemShut
  {NoStop}%
\bibitem [{\citenamefont {Takami}\ \emph {et~al.}(2015)\citenamefont {Takami},
  \citenamefont {Rezzolla},\ and\ \citenamefont {Baiotti}}]{Takami:2014tva}%
  \BibitemOpen
  \bibfield  {author} {\bibinfo {author} {\bibfnamefont {K.}~\bibnamefont
  {Takami}}, \bibinfo {author} {\bibfnamefont {L.}~\bibnamefont {Rezzolla}}, \
  and\ \bibinfo {author} {\bibfnamefont {L.}~\bibnamefont {Baiotti}},\ }\href
  {\doibase 10.1103/PhysRevD.91.064001} {\bibfield  {journal} {\bibinfo
  {journal} {Phys. Rev.}\ }\textbf {\bibinfo {volume} {D91}},\ \bibinfo {pages}
  {064001} (\bibinfo {year} {2015})},\ \Eprint {http://arxiv.org/abs/1412.3240}
  {arXiv:1412.3240 [gr-qc]} \BibitemShut {NoStop}%
\bibitem [{\citenamefont {Damour}(2001)}]{Damour:2001tu}%
  \BibitemOpen
  \bibfield  {author} {\bibinfo {author} {\bibfnamefont {T.}~\bibnamefont
  {Damour}},\ }\href {\doibase 10.1103/PhysRevD.64.124013} {\bibfield
  {journal} {\bibinfo  {journal} {Phys. Rev.}\ }\textbf {\bibinfo {volume}
  {D64}},\ \bibinfo {pages} {124013} (\bibinfo {year} {2001})},\ \Eprint
  {http://arxiv.org/abs/gr-qc/0103018} {arXiv:gr-qc/0103018 [gr-qc]}
  \BibitemShut {NoStop}%
\bibitem [{\citenamefont {Blanchet}(2014)}]{Blanchet:2013haa}%
  \BibitemOpen
  \bibfield  {author} {\bibinfo {author} {\bibfnamefont {L.}~\bibnamefont
  {Blanchet}},\ }\href {\doibase 10.12942/lrr-2014-2} {\bibfield  {journal}
  {\bibinfo  {journal} {Living Rev. Rel.}\ }\textbf {\bibinfo {volume} {17}},\
  \bibinfo {pages} {2} (\bibinfo {year} {2014})},\ \Eprint
  {http://arxiv.org/abs/1310.1528} {arXiv:1310.1528 [gr-qc]} \BibitemShut
  {NoStop}%
\bibitem [{\citenamefont {Damour}\ \emph {et~al.}(2012)\citenamefont {Damour},
  \citenamefont {Nagar},\ and\ \citenamefont {Villain}}]{PhysRevD.85.123007}%
  \BibitemOpen
  \bibfield  {author} {\bibinfo {author} {\bibfnamefont {T.}~\bibnamefont
  {Damour}}, \bibinfo {author} {\bibfnamefont {A.}~\bibnamefont {Nagar}}, \
  and\ \bibinfo {author} {\bibfnamefont {L.}~\bibnamefont {Villain}},\ }\href
  {\doibase 10.1103/PhysRevD.85.123007} {\bibfield  {journal} {\bibinfo
  {journal} {Phys. Rev. D}\ }\textbf {\bibinfo {volume} {85}},\ \bibinfo
  {pages} {123007} (\bibinfo {year} {2012})}\BibitemShut {NoStop}%
\bibitem [{\citenamefont {Dietrich}\ \emph {et~al.}(2019)\citenamefont
  {Dietrich}, \citenamefont {Samajdar}, \citenamefont {Khan}, \citenamefont
  {Johnson-McDaniel}, \citenamefont {Dudi},\ and\ \citenamefont
  {Tichy}}]{Dietrich:2019kaq}%
  \BibitemOpen
  \bibfield  {author} {\bibinfo {author} {\bibfnamefont {T.}~\bibnamefont
  {Dietrich}}, \bibinfo {author} {\bibfnamefont {A.}~\bibnamefont {Samajdar}},
  \bibinfo {author} {\bibfnamefont {S.}~\bibnamefont {Khan}}, \bibinfo {author}
  {\bibfnamefont {N.~K.}\ \bibnamefont {Johnson-McDaniel}}, \bibinfo {author}
  {\bibfnamefont {R.}~\bibnamefont {Dudi}}, \ and\ \bibinfo {author}
  {\bibfnamefont {W.}~\bibnamefont {Tichy}},\ }\href@noop {} {\  (\bibinfo
  {year} {2019})},\ \Eprint {http://arxiv.org/abs/1905.06011} {arXiv:1905.06011
  [gr-qc]} \BibitemShut {NoStop}%
\bibitem [{\citenamefont {Khan}\ \emph {et~al.}(2016)\citenamefont {Khan},
  \citenamefont {Husa}, \citenamefont {Hannam}, \citenamefont {Ohme},
  \citenamefont {Pürrer}, \citenamefont {Jiménez~Forteza},\ and\
  \citenamefont {Bohé}}]{Khan:2015jqa}%
  \BibitemOpen
  \bibfield  {author} {\bibinfo {author} {\bibfnamefont {S.}~\bibnamefont
  {Khan}}, \bibinfo {author} {\bibfnamefont {S.}~\bibnamefont {Husa}}, \bibinfo
  {author} {\bibfnamefont {M.}~\bibnamefont {Hannam}}, \bibinfo {author}
  {\bibfnamefont {F.}~\bibnamefont {Ohme}}, \bibinfo {author} {\bibfnamefont
  {M.}~\bibnamefont {Pürrer}}, \bibinfo {author} {\bibfnamefont
  {X.}~\bibnamefont {Jiménez~Forteza}}, \ and\ \bibinfo {author}
  {\bibfnamefont {A.}~\bibnamefont {Bohé}},\ }\href {\doibase
  10.1103/PhysRevD.93.044007} {\bibfield  {journal} {\bibinfo  {journal} {Phys.
  Rev.}\ }\textbf {\bibinfo {volume} {D93}},\ \bibinfo {pages} {044007}
  (\bibinfo {year} {2016})},\ \Eprint {http://arxiv.org/abs/1508.07253}
  {arXiv:1508.07253 [gr-qc]} \BibitemShut {NoStop}%
\bibitem [{\citenamefont {Biwer}\ \emph {et~al.}(2019)\citenamefont {Biwer},
  \citenamefont {Capano}, \citenamefont {De}, \citenamefont {Cabero},
  \citenamefont {Brown}, \citenamefont {Nitz},\ and\ \citenamefont
  {Raymond}}]{Biwer:2018osg}%
  \BibitemOpen
  \bibfield  {author} {\bibinfo {author} {\bibfnamefont {C.~M.}\ \bibnamefont
  {Biwer}}, \bibinfo {author} {\bibfnamefont {C.~D.}\ \bibnamefont {Capano}},
  \bibinfo {author} {\bibfnamefont {S.}~\bibnamefont {De}}, \bibinfo {author}
  {\bibfnamefont {M.}~\bibnamefont {Cabero}}, \bibinfo {author} {\bibfnamefont
  {D.~A.}\ \bibnamefont {Brown}}, \bibinfo {author} {\bibfnamefont {A.~H.}\
  \bibnamefont {Nitz}}, \ and\ \bibinfo {author} {\bibfnamefont
  {V.}~\bibnamefont {Raymond}},\ }\href {\doibase 10.1088/1538-3873/aaef0b}
  {\bibfield  {journal} {\bibinfo  {journal} {Publ. Astron. Soc. Pac.}\
  }\textbf {\bibinfo {volume} {131}},\ \bibinfo {pages} {024503} (\bibinfo
  {year} {2019})},\ \Eprint {http://arxiv.org/abs/1807.10312} {arXiv:1807.10312
  [astro-ph.IM]} \BibitemShut {NoStop}%
\end{thebibliography}%
